\documentclass[aps,prc,twocolumn,superscriptaddress,floatfix]{revtex4-1}
\usepackage[english]{babel}
\usepackage[T1]{fontenc}
\usepackage[colorlinks=true,linkcolor=blue,citecolor=blue,urlcolor=blue]{hyperref}
\usepackage{bm,bbm,amssymb,amsmath}
\usepackage{graphicx}
\usepackage{multirow,booktabs}
\usepackage{isotope}
\usepackage{physics}
\usepackage[capitalize]{cleveref}

\begin{document}

\title{Auxiliary field diffusion Monte Carlo calculations of light and medium-mass nuclei with local chiral interactions}

\author{D. Lonardoni}
\affiliation{Facility for Rare Isotope Beams, Michigan State University, East Lansing, MI 48824, USA}
\affiliation{Theoretical Division, Los Alamos National Laboratory, Los Alamos, New Mexico 87545, USA}

\author{S. Gandolfi}
\affiliation{Theoretical Division, Los Alamos National Laboratory, Los Alamos, New Mexico 87545, USA}

\author{J.~E. Lynn}
\affiliation{Institut f\"ur Kernphysik, Technische Universit\"at Darmstadt, 64289 Darmstadt, Germany}
\affiliation{ExtreMe Matter Institute EMMI, GSI Helmholtzzentrum f\"ur Schwerionenforschung GmbH, 64291 Darmstadt, Germany}

\author{C. Petrie}
\affiliation{Department of Physics, Arizona State University, Tempe, Arizona 85287, USA}

\author{J. Carlson}
\affiliation{Theoretical Division, Los Alamos National Laboratory, Los Alamos, New Mexico 87545, USA}

\author{K.~E. Schmidt}
\affiliation{Department of Physics, Arizona State University, Tempe, Arizona 85287, USA}

\author{A. Schwenk}
\affiliation{Institut f\"ur Kernphysik, Technische Universit\"at Darmstadt, 64289 Darmstadt, Germany}
\affiliation{ExtreMe Matter Institute EMMI, GSI Helmholtzzentrum f\"ur Schwerionenforschung GmbH, 64291 Darmstadt, Germany}
\affiliation{Max-Planck-Institut f\"ur Kernphysik, Saupfercheckweg 1, 69117 Heidelberg, Germany}

\begin{abstract}
Quantum Monte Carlo methods have recently been employed to study properties
of nuclei and infinite matter using local chiral effective field theory 
interactions. In this work, we present a detailed description of the
auxiliary field diffusion Monte Carlo algorithm for nuclei in combination 
with local chiral two- and three-nucleon interactions up to 
next-to-next-to-leading order. We show results for the binding energy, 
charge radius, charge form factor, and Coulomb sum rule in nuclei with
$3\le A\le16$. Particular attention is devoted to the effect 
of different operator structures in the three-body force for different cutoffs. 
The outcomes suggest that local chiral interactions fit to few-body observables
give a very good description of the ground-state properties of nuclei up to
\isotope[16]{O}, with the exception of one fit for the softer cutoff
which predicts overbinding in larger nuclei.
\end{abstract}

\maketitle

\section{Introduction}
\label{sec:intro}
The solution of the many-body Schr\"odinger equation describing a system of  
interacting baryons is challenging because of the nonperturbative 
nature and the strong spin/isospin-dependence of realistic nuclear interactions. 
Quantum Monte Carlo (QMC) methods provide a powerful tool to tackle the nuclear 
many-body problem in a nonperturbative fashion. They have been proven to be 
remarkably successful in describing the properties of strongly correlated 
fermions in a large variety of physical conditions~\cite{Carlson:2015}. 

Historically, QMC methods have made use of phenomenological nuclear interactions, 
such as the Argonne $v_{18}$ (AV18) nucleon-nucleon $(N\!N)$ potential combined
with Urbana/Illinois models for the three-nucleon $(3N)$ forces~\cite{Carlson:2015}.
By construction, these potentials are nearly local, meaning that the dominant parts 
of the interaction depend only on the relative distance, spin, and isospin of the
two interacting nucleons, and not upon their momenta.
This feature has been one of the keys to success
for the application of QMC algorithms to the study of nuclear systems. 
Green's function Monte Carlo (GFMC) and auxiliary field diffusion Monte Carlo (AFDMC) 
methods have employed these phenomenological potentials to accurately calculate properties 
of nuclei, neutron drops, and neutron-star 
matter~\cite{Carlson:2015,Gandolfi:2011,Gandolfi:2012,Maris:2013,Gandolfi:2014,Gandolfi:2014_epja,Buraczynski:2016,Buraczynski:2017}.
Despite the large success of such models, phenomenological interactions are not free from
drawbacks. They do not provide a systematic way to estimate theoretical uncertainties, 
and it is not clear how to improve their quality. In addition, some models of the 
$3N$ force provide a too soft equation of state of neutron matter~\cite{Sarsa:2003,Maris:2013},
with the consequence that the predicted neutron-star maximum mass is not compatible
with the observation of heavy neutron stars~\cite{Demorest:2010,Antoniadis:2013}.

An alternative approach to nuclear interactions that overcomes the limitations
of the phenomenological models is provided by chiral effective field theory 
(EFT)~\cite{Epelbaum:2009,Machleidt:2011}.
In chiral EFT, nuclear interactions are systematically derived in connection with 
the underlying theory of the strong interaction, by writing down the most general 
Lagrangian consistent with the symmetries of low-energy quantum chromodynamics
(QCD) in terms of the relevant degrees of freedom at low energies; nucleons and pions.
A power-counting scheme is then chosen to order the resulting contributions according 
to their importance. The result is a low-energy EFT according
to which nuclear forces are given in an expansion in the ratio of a 
soft scale (the pion mass or a typical momentum scale in the nucleus) to a
hard scale (the chiral breakdown scale). The long-range part of the potential
is given by pion-exchange contributions that are determined by the chiral symmetry
of QCD and low-energy experimental data for the pion-nucleon system.
The short-range terms are instead characterized by contact
interactions described by so-called low-energy constants (LECs), 
which are fit to reproduce experimental data ($N\!N$ scattering data for 
the two-body part of the interaction, and few- and/or many-body observables for 
the many-body components). 
Among the advantages of such an 
expansion, compared to traditional approaches, are the ability to 
systematically improve the quality of the interaction order by order, 
the possibility to estimate theoretical uncertainties, the fact that
many-body forces arise naturally, and that electroweak currents can 
be derived consistently.

In the last decade, intense efforts have been devoted to the 
development of chiral EFT interactions, as shown by the 
availability of different potentials in the 
literature~\cite{Epelbaum:2009,Machleidt:2011,Epelbaum:2015,Ekstrom:2015,Carlsson:2016,Entem:2017,Ekstrom:2017,Reinert:2017}, 
typically written in momentum 
space. It is only in recent
years that chiral EFT interactions have been formulated equivalently 
in coordinate space. New potentials are now available, including 
next-to-next-to-leading-order (N$^2$LO) local 
interactions~\cite{Gezerlis:2013,Gezerlis:2014}, supplemented by consistent
$3N$ potentials~\cite{Tews:2016,Lynn:2016}, as
well as chiral interactions with explicit delta degrees of
freedom~\cite{Piarulli:2015,Piarulli:2016,Piarulli:2018}.

Local chiral interactions up to N$^2$LO can be written using the same operator
structure as the phenomenological potentials, providing for the first time
the opportunity to combine EFT-derived interactions and accurate QMC methods.
The GFMC method has been used to study the ground state of light nuclei employing 
local chiral 
interactions~\cite{Gezerlis:2013,Gezerlis:2014,Lynn:2014,Tews:2016,Lynn:2016,Piarulli:2016,Lynn:2017,Chen:2017,Piarulli:2018}.
The same potentials have been used in AFDMC calculations of pure neutron systems, 
ranging from few-body systems~\cite{Klos:2016,Zhao:2016,Gandolfi:2017} to pure
neutron matter~\cite{Gezerlis:2013,Gezerlis:2014,Tews:2016}.
More recently, the first AFDMC study of $p$-shell nuclei employing local
chiral interactions has been reported~\cite{Lonardoni:2017afdmc}.
In this work we provide a comprehensive description of the AFDMC algorithm 
for the study of ground-state properties of light and medium-mass nuclei 
employing local chiral interactions at N$^2$LO, extending
the findings of Ref.~\cite{Lonardoni:2017afdmc}.

The structure of this paper is as follows. 
In~\cref{sec:ham} we introduce the nuclear Hamiltonian employed in this work.
In~\cref{sec:vmc,sec:afdmc} we review the main features of the employed QMC methods.
\Cref{sec:wf} is devoted to the description of the employed trial wave functions. 
In~\cref{sec:res} we present our results for nuclei
with $3\le A\le 16$. 
Finally, we give a summary in~\cref{sec:summ}.

\section{Hamiltonian}
\label{sec:ham}
Nuclei are described as a collection of point-like particles of mass $m_N$ interacting 
via two- and three-body potentials according to the nonrelativistic Hamiltonian
\begin{align}
	H=-\frac{\hbar^2}{2m_N}\sum_i \nabla_i^2+\sum_{i<j}v_{ij}+\sum_{i<j<k}V_{ijk} ,
\end{align}
where the two-body interaction $v_{ij}$ also includes the Coulomb force.

In QMC calculations, it is convenient to express the interactions
in terms of radial functions multiplying spin and isospin operators. The commonly 
employed Argonne $v_8'$ (AV8$'$) potential~\cite{Wiringa:2002}, as well as the two-body
part of the recently developed local chiral interactions~\cite{Gezerlis:2013}, 
can be expressed as:
\begin{align}
	v_{ij} = \sum_{p=1}^8 v_p(r_{ij}) \mathcal O_{ij}^{p},
	\label{eq:v_ij}
\end{align}
with
\begin{align}
	\mathcal O_{ij}^{p=1,8} = \big[\mathbbm 1,\bm\sigma_i\cdot\bm\sigma_j,S_{ij},\vb{L}\cdot\vb{S}\big]
	\otimes\big[\mathbbm 1,\bm\tau_i\cdot\bm\tau_j\big],
\end{align}
where
\begin{align}
	S_{ij}=3\,\bm\sigma_i\cdot\hat{\vb{r}}_{ij}\,\bm\sigma_j\cdot\hat{\vb{r}}_{ij}-\bm\sigma_i\cdot\bm\sigma_j,
\end{align}
is the tensor operator, and
\begin{align}
	\vb{L}&=\frac{1}{2i}(\vb{r}_i-\vb{r}_j)\times(\bm\nabla_i-\bm\nabla_j), \label{eq:L} \\
	\vb{S}&=\frac{1}{2}(\bm\sigma_i+\bm\sigma_j), \label{eq:S}
\end{align}
are the relative angular momentum and the total spin of the pair $ij$, respectively.
The radial functions of~\cref{eq:v_ij} are fitted to $N\!N$ scattering data.
At N$^2$LO, the operator structure of the local chiral interactions is the same as 
above, with the only exception that the $\vb{L}\cdot\vb{S}\,\bm\tau_i\cdot\bm\tau_j$
term is not present at N$^2$LO (see Ref.~\cite{Gezerlis:2014} for more details).

The three-body force $V_{ijk}$ is written as a sum of contributions coming from two-pion
exchange (TPE), plus shorter-range terms. In the case of local chiral interactions at N$^2$LO, $P$- and $S$-wave
TPE contributions are included, and they are characterized by the same LECs involved in the two-body sector. 
The shorter-range part of the $3N$ force is instead parametrized by two contact terms, 
the LECs of which have been fit to the alpha particle binding energy and to the spin-orbit splitting in 
the neutron-$\alpha$ $P$-wave phase shifts (see Refs.~\cite{Lynn:2016,Lynn:2017} for more details).

The chiral $3N$ interaction at N$^2$LO can be conveniently written as
\begin{align}
V=V_a^{2\pi,P}+V_c^{2\pi,P}+V^{2\pi,S}+V_D+V_E,
\label{eq:v_ijk}
\end{align}
where the first three terms correspond to the TPE diagrams in $P$ and $S$ waves (Eqs. (A.1b),
(A.1c) and (A.1a) of Ref.~\cite{Lynn:2017}, respectively). The subscripts $a$ and $c$ refer 
to the operator structure of such contributions, which can be written in terms of anticommutators 
or commutators, respectively. $V_D$ and $V_E$ involve contact terms. In this work we employ
the form (A.2b) of Ref.~\cite{Lynn:2017} for $V_D$, and we consider two choices for $V_E$, 
namely $E\tau$ and $E\mathbbm1$ (Eqs.~(A.3a) and (A.3b) of Ref.~\cite{Lynn:2017}).

By defining the following quantities:
\begin{align}
\delta_{R_0}(r)&=\frac{n}{4\pi R_0^3\Gamma(3/n)}\,e^{-(r/R_0)^n}
\nonumber, \\
T(r)&=\left(1+\frac{3}{m_\pi r}+\frac{3}{m_\pi^2r^2}\right)\frac{e^{-m_\pi r}}{m_\pi r} T_c(r)
\nonumber, \\
Y(r)&=\frac{e^{-m_\pi r}}{m_\pi r}\,Y_c(r) 
\nonumber, \\
Z(r)&=\frac{m_\pi r}{3}\Big(Y(r)-T(r)\Big)
\nonumber, \\
Y_c(r)&=1-e^{-(r/R_0)^n}
\nonumber, \\
T_c(r)&=\left(1-e^{-(r/R_0)^n}\right)^{n_t}
\nonumber, \\
X_{i\alpha j\beta}&=(3\,\delta_{\alpha\gamma}\,\hat{r}_{ij}^\gamma\;\delta_{\beta\mu}\,\hat{r}_{ij}^\mu-\delta_{\alpha\beta})\,T(r_{ij})
+\delta_{\alpha\beta}\,Y(r_{ij}) 
\nonumber, \\
{\cal X}_{i\alpha j\beta}&=X_{i\alpha j\beta}(\vb{r}_{ij})-\delta_{\alpha\beta}\frac{4\pi}{m_\pi^3}\delta_{R_0}(r_{ij})
\nonumber, \\
{\cal Z}_{ij\alpha}&=Z(r_{ij})\,\delta_{\alpha\gamma}\,\hat{r}_{ij}^\gamma,
\end{align}
we can recast the contributions of~\cref{eq:v_ijk} in a form that is suitable for QMC calculations:
\begin{widetext}
\begin{align}
V_a^{2\pi,P}&=A_a^{2\pi,P}\sum_{i<j<k}\sum_{\rm cyc}\Big\{\bm{\tau}_i\cdot\bm{\tau}_k,\bm{\tau}_j\cdot\bm{\tau}_k\Big\}
\Big\{\sigma_i^\alpha\sigma_k^\gamma,\sigma_k^\mu\sigma_j^\beta\Big\}
{\cal X}_{i\alpha k\gamma}\,{\cal X}_{k\mu j\beta}
\nonumber \\
&=4\,A_a^{2\pi,P}\sum_{i<j}\bm{\tau}_i\cdot\bm{\tau}_j\,\sigma_i^\alpha\sigma_j^\beta
\sum_{k\neq i,j}{\cal X}_{i\alpha k\gamma}\,{\cal X}_{k\mu j\beta} 
\nonumber \\
&=4\,A_a^{2\pi,P}\sum_{i<j}\bm{\tau}_i\cdot\bm{\tau}_j\,\sigma_i^\alpha\sigma_j^\beta
\sum_{k\neq i,j}\left(X_{i\alpha k\gamma}-\delta_{\alpha\gamma}\frac{4\pi}{m_\pi^3}\delta_{R_0}(r_{ik})\right)
\left(X_{k\mu j\beta}-\delta_{\mu\beta}\frac{4\pi}{m_\pi^3}\delta_{R_0}(r_{kj})\right)
\nonumber \\
&=V_a^{XX}+V_a^{X\delta}+V_a^{\delta\delta} ,
\label{eq:anti}
\\
V_c^{2\pi,P}&=A_c^{2\pi,P}\sum_{i<j<k}\sum_{\rm cyc}\Big[\bm{\tau}_i\cdot\bm{\tau}_k,\bm{\tau}_j\cdot\bm{\tau}_k\Big]
\Big[\sigma_i^\alpha\sigma_k^\gamma,\sigma_k^\mu\sigma_j^\beta\Big]
{\cal X}_{i\alpha k\gamma}\,{\cal X}_{k\mu j\beta}
\nonumber \\
&=A_c^{2\pi,P}\sum_{i<j<k}\sum_{\rm cyc}\Big[\bm{\tau}_i\cdot\bm{\tau}_k,\bm{\tau}_j\cdot\bm{\tau}_k\Big]
\Big[\sigma_i^\alpha\sigma_k^\gamma,\sigma_k^\mu\sigma_j^\beta\Big]
\left(X_{i\alpha k\gamma}-\delta_{\alpha\gamma}\frac{4\pi}{m_\pi^3}\delta_{R_0}(r_{ik})\right)
\left(X_{k\mu j\beta}-\delta_{\mu\beta}\frac{4\pi}{m_\pi^3}\delta_{R_0}(r_{kj})\right) 
\nonumber \\
&=V_c^{XX}+V_c^{X\delta}+V_c^{\delta\delta} ,
\label{eq:comm}
\\
V^{2\pi,S}&=A^{2\pi,S}\sum_{i<j<k}\sum_{\rm cyc}
\bm{\tau}_i\cdot\bm{\tau}_j\,\sigma_i^\alpha\sigma_j^\beta\,{\cal Z}_{ik\alpha}\,{\cal Z}_{jk\alpha}
\nonumber \\
&=A^{2\pi,S}\sum_{i<j}\bm{\tau}_i\cdot\bm{\tau}_j\,\sigma_i^\alpha\sigma_j^\beta
\sum_{k\neq i,j}{\cal Z}_{ik\alpha}\,{\cal Z}_{jk\alpha} ,
\label{eq:tm}
\\
V_D&=A_D\sum_{i<j}\bm{\tau}_i\cdot\bm{\tau}_j\,\sigma_i^\alpha\sigma_j^\beta
\sum_{k\neq i,j}{\cal X}_{i\alpha j\beta}\Big[\delta_{R_0}(r_{ik})+\delta_{R_0}(r_{jk})\Big]  
\nonumber \\
&=A_D\sum_{i<j}\bm{\tau}_i\cdot\bm{\tau}_j\,\sigma_i^\alpha\sigma_j^\beta
\sum_{k\neq i,j}\left(X_{i\alpha j\beta}-\delta_{\alpha\beta}\frac{4\pi}{m_\pi^3}\delta_{R_0}(r_{ij})\right) \Big[\delta_{R_0}(r_{ik})+\delta_{R_0}(r_{jk})\Big]  
\nonumber \\
&=V_D^{X\delta}+V_D^{\delta\delta} ,
\label{eq:vd}
\\
V_E&=A_E\sum_{i<j}\bm{\tau}_i\cdot\bm{\tau}_j\sum_{k\neq i,j}\delta_{R_0}(r_{ik})\delta_{R_0}(r_{jk}) ,
\label{eq:ve}
\end{align}
\end{widetext}
where the sum over the coordinate projections (Greek indices) is implicit.
\Cref{eq:ve} is the expression for the $E\tau$ parametrization of the contact term $V_E$.
The $E\mathbbm1$ form is recovered by setting $\bm{\tau}_i\cdot\bm{\tau}_j=\mathbbm1$.
For the local chiral interactions at N$^2$LO, we have
\begin{align}
A_a^{2\pi,P}&=\frac{1}{2}\left(\frac{g_A}{f_\pi^2}\right)^2\left(\frac{1}{4\pi}\right)^2
\frac{m_\pi^6}{9}c_3 \nonumber, \\
A_c^{2\pi,P}&=-\frac{c_4}{2c_3}A_a^{2\pi,P} \nonumber, \\
A^{2\pi,S}&=\left(\frac{g_A}{2f_\pi}\right)^2\left(\frac{m_\pi}{4\pi}\right)^2\frac{4m_\pi^6}{f_\pi^2}c_1
\nonumber, \\
A_D&=\frac{m_\pi^3}{12\pi}\frac{g_A}{8f_\pi^2}\frac{1}{f_\pi^2\Lambda_\chi}c_D
\nonumber, \\
A_E&=\frac{c_E}{f_\pi^4\Lambda_\chi},
\label{eq:lecs}
\end{align}
where $g_A=1.267$ is the axial-vector coupling constant, $f_\pi=92.4\,\rm MeV$ 
is the pion decay constant, $m_\pi=138.03\,\rm MeV$ is the averaged pion mass, 
$\Lambda_\chi$ is taken to be a heavy meson scale $\Lambda_\chi=700\,\rm MeV$, and
$c_1,\,c_3,\,c_4,\,c_D,\,c_E$ are the LECs.
Note that, using these definitions, the structure of the phenomenological Urbana IX (UIX) model
is recovered by imposing $\delta_{R_0}(r)=0$, $n=2$, $n_t=2$, and $A_c^{2\pi,P}=\frac{1}{4}A_a^{2\pi,P}$
as well as $A_D=A_E=0$.
Finally, we consider coordinate-space cutoffs $R_0=1.0\,\rm fm$ and $R_0=1.2\,\rm fm$, 
approximately corresponding to cutoffs in momentum space of $500\,\rm MeV$ and $400\,\rm MeV$~\cite{Lynn:2017}, 
note however also Ref.~\cite{Hoppe:2017}.

\section{Review of the VMC method}
\label{sec:vmc}
In the variational Monte Carlo (VMC) method, given a trial wave function $\Psi_T$,
the expectation value of the Hamiltonian $H$ is given by
\begin{align}
\!\!\!E_0\leq\langle H\rangle=\frac{\langle\Psi_T|H|\Psi_T\rangle}{\langle\Psi_T|\Psi_T\rangle}
=\frac{\displaystyle\int\!dR\,\Psi_T^*(R)H\Psi_T(R)}
{\displaystyle\int\!dR\,\Psi_T^*(R)\Psi_T(R)} ,
\label{eq:<h>}
\end{align}
where $R=\{\vb{r}_1,\dots,\vb{r}_A\}$ are the coordinates of the particles, and there is an 
implicit sum over all the particle spin and isospin states. 
$E_0$ is the energy of the true ground state with the same quantum numbers as $\Psi_T$, 
and the leftmost equality in the above relation is valid only if the wave function is the exact
ground-state wave function $\Psi_0$. In the VMC method, one typically minimizes the energy expectation 
value of~\cref{eq:<h>} with respect to changes in the variational parameters, in order to 
obtain $\Psi_T$ as close as possible to $\Psi_0$. 

The integral of~\cref{eq:<h>} can be rewritten as
\begin{align}
\langle H\rangle=\frac{\displaystyle\int dR\,P(R) \frac{H\Psi_T(R)}{\Psi_T(R)}}{\displaystyle\int dR\,P(R)} ,
\end{align}
where $P(R)=|\Psi_T(R)|^2$ can be interpreted as a probability distribution of points $R$ in a $3A$-dimensional space.
The above multidimensional integral can be solved using Monte Carlo sampling.
In practice, a number of configurations $R_i$ are sampled using the Metropolis algorithm~\cite{Metropolis:1953}, 
and the local energy of the system is calculated as:
\begin{align}
\langle E\rangle=\frac{1}{A}\sum_{i=1}^A \frac{\langle R_i|H|\Psi_T\rangle}{\langle R_i|\Psi_T\rangle} ,
\end{align}
where $\langle R|\Psi_T\rangle=\Psi_T(R)$.
More details on the sampling procedure and on the calculation of statistical errors can be found, 
e.g., in Ref.~\cite{Ceperley:1995}.

For spin/isospin-dependent interactions the generalization of~\cref{eq:<h>} is straightforward:
\begin{align}
\langle H\rangle=\frac{\displaystyle\int dR\,\sum_{S,S'}\Psi_T^*(R,S')H_{S,S'}\Psi_T(R,S)}
{\displaystyle\int dR\,\sum_S|\Psi_T(R,S)|^2} ,
\end{align}
where now the wave function also depends upon spin and isospin states $S=\{s_1,\dots,s_A\}$, and
\begin{align}
H_{S,S'}=\langle S'|S\rangle\left[-\frac{\hbar^2}{2m}\sum_i\nabla_i^2\right]+\langle RS'|V|RS\rangle .
\end{align}
In this case, the VMC method can be implemented by either explicitly summing over all the spin and isospin states
\begin{align}
&\langle H\rangle=\displaystyle\int dR\,E_L(R)P(R) ,\nonumber \\
&P(R)=\frac{\sum_S|\Psi_T(R,S)|^2}{\displaystyle\int dR\,\sum_S|\Psi_T(R,S)|^2} ,\nonumber \\
&E_L(R)=\frac{\sum_{S,S'}\Psi_T^*(R,S')H_{S,S'}\Psi_T(R,S)}{\sum_S|\Psi_T(R,S)|^2} ,
\end{align}
or by sampling the spin and isospin states:
\begin{align}
&\langle H\rangle=\displaystyle\int dR\,\sum_S\,E_L(R,S)P(R,S) ,\nonumber \\
&P(R,S)=\frac{|\Psi_T(R,S)|^2}{\displaystyle\int dR\,|\Psi_T(R,S)|^2} ,\nonumber \\
&E_L(R,S)=\frac{\sum_{S'}\Psi_T^*(R,S')H_{S,S'}\Psi_T(R,S)}{|\Psi_T(R,S)|^2} .
\end{align}
The Metropolis algorithm can then be used to sample either $R$ from $P(R)$ in the former case, 
or $R$ and $S$ from $P(R,S)$ in the latter case.

\section{Review of the AFDMC method}
\label{sec:afdmc}
Diffusion Monte Carlo (DMC) methods are used to project out the ground state
with a particular set of quantum numbers. The starting point is a trial wave
function $|\Psi_T\rangle$, typically the result of a VMC minimization, that is
propagated in imaginary time $\tau$:
\begin{align}
|\Psi_0\rangle\propto \lim_{\tau\rightarrow\infty} e^{-(H-E_T)\tau}|\Psi_T\rangle ,
\end{align}
where $E_T$ is a parameter that controls the normalization. 
For spin/isospin-independent interactions, the object to be propagated 
is given by the overlap between the wave function and a set of configurations in 
coordinate space $\langle R|\Psi_T\rangle=\Psi_T(R)$. 
By using the completeness relation $\int dR |R\rangle\langle R|=\mathbbm1$,
we can write the propagation in imaginary time as
\begin{align}
\langle R'|\Psi(\tau)\rangle=\displaystyle\int dR\,G(R',R,\tau)\,\langle R|\Psi_T(0)\rangle ,
\label{eq:imtimeprop}
\end{align}
where the propagator (or Green's function) $G$ is defined as the matrix element between
the two points $R$ and $R'$ in the volume
\begin{align}
G(R',R,\tau)=\langle R'|e^{-(H-E_T)\tau}|R\rangle,
\end{align}
and $\langle R'|\Psi(\tau)\rangle$ approaches the true ground-state 
for large imaginary time.

In practice, it is not possible to directly compute
the propagator $G(R',R,\tau)$. 
However, one can use the short-time propagator $G(R',R,d\tau)$:
\begin{align}
\langle R'|\Psi(\tau)\rangle=
\int dR_n\,dR_{n-1}\ldots\,dR_1\,dR\,G(R',R_n,\delta\tau) \nonumber \\
\times G(R_{n-1},R_{n-2},\delta\tau)\ldots\,G(R_1,R,\delta\tau)
\langle R|\Psi_T(0)\rangle,
\end{align}
and then employ Monte Carlo techniques to sample the paths $R_i$ in the
imaginary-time evolution. The method is accurate for small values of 
the time step $\delta\tau$, and the exact result 
can be determined by using different values of $\delta\tau$ 
and extrapolating to $\delta\tau\to0$.

By using the Trotter formula~\cite{Trotter:1959} to order $d\tau^3$,
the short-time propagator can be approximated with:
\begin{align}
G(R',R,\delta\tau)&\equiv\langle R'|e^{-(H-E_T)\delta\tau}|R\rangle \nonumber \\
&\approx \langle R'|e^{-(V-E_T)\frac{\delta\tau}{2}}e^{-T\delta\tau}e^{-(V-E_T)\frac{\delta\tau}{2}}
|R\rangle ,
\end{align}
where $T$ is the nonrelativistic kinetic energy, and $V$ is the employed potential.
The propagator for the kinetic energy alone corresponds to the free-particle propagator:
\begin{align} 
G_0(R',R)&=\langle R'|e^{-T\delta\tau}|R\rangle\nonumber \\
&=\left(\frac{m}{2\pi\hbar^2\delta\tau}\right)^{\frac{3A}{2}}e^{-\frac{m(R-R')^2}{2\hbar^2\delta\tau}} ,
\label{eq:g0}
\end{align}
which yields a Gaussian diffusion for the paths in coordinate space, with
$\sigma^2 = 4 \frac{\hbar^2}{2m} \delta\tau$.
The propagator for spin/isospin-independent potentials is simply given by: 
\begin{align}
\langle R'|e^{-(V-E_T)\delta\tau}|R\rangle \approx
\prod_{i<j}e^{-[V(r_{ij})-E_T]\delta\tau}\,\delta(R-R') ,
\label{eq:p_vr}
\end{align}
where each pair interaction can be simply evaluated as a function of the coordinates of the 
system, and the energy $E_T$ results in a normalization factor. 
Note that the addition of spin/isospin-independent three- and many-body interactions
is straightforward.

For spin/isospin-dependent interactions, the propagation of the potential becomes more complicated.
In general, this is because quadratic operators like $\bm\sigma_i\cdot\bm\sigma_j$
generate amplitudes along the singlet and the triplet states of a pair. 
The propagator of~\cref{eq:p_vr} generalizes in this case to
\begin{align}
\langle R'|e^{-(V-E_T)\delta\tau}|R\rangle  \rightarrow
\langle R'S'|e^{-(V-E_T)\delta\tau}|RS\rangle \nonumber \\
 \approx \langle S'|\prod_{i<j}e^{-(V(r_{ij})-E_T)\delta\tau}|S\rangle\,\delta(R-R') ,
\label{eq:p_vrs}
\end{align}
where now the matrix $\exp[(-(V-E_T)\delta\tau)]$ is not diagonal in the spin of each pair.
One possible strategy to compute the propagator of~\cref{eq:p_vrs} is to include all the  
spin and isospin states in the trial wave function, as is done in GFMC calculations~\cite{Carlson:2015}.  
This, however, implies a number of wave-function components proportional to $2^A$, 
which currently limits GFMC calculations to $A=12$.

The idea of the AFDMC method is to start from a trial wave function
whose computational cost is polynomial with $A$, rather than exponential.
Such a wave function can be written in the single-particle representation:
\begin{align}
\langle S|\Psi\rangle\propto\xi_{\alpha_1}(s_1)\,\xi_{\alpha_2}(s_2)\dots\,\xi_{\alpha_A}(s_A) ,
\label{eq:wf_s}
\end{align}
where $\xi_{\alpha_i}(s_i)$ are functions of the spinor $s_i$ with state $\alpha_i$. 
In the above expression, the radial orbitals are omitted for simplicity, 
and the antisymmetrization is trivial.

A quadratic operator in the spin acting on the wave function above generates two different amplitudes:
\begin{align}
\langle S|\bm\sigma_1\cdot\bm\sigma_2|\Psi\rangle & =\langle S|2\,\mathcal P_{12}^\sigma-\mathbbm1|\Psi\rangle \nonumber \\
& =2\,\xi_{\alpha_1}(s_2)\,\xi_{\alpha_2}(s_1)\,\xi_{\alpha_3}(s_3)\dots\,\xi_{\alpha_A}(s_A) \nonumber \\
& \quad\,-\xi_{\alpha_1}(s_1)\,\xi_{\alpha_2}(s_2)\,\xi_{\alpha_3}(s_3)\dots\,\xi_{\alpha_A}(s_A) \nonumber \\
& =\langle S'|\Psi\rangle+\langle S''|\Psi\rangle .
\end{align}
In general, the action of all pairwise spin/isospin operators (or propagators) generates $2^A\binom{A}{Z}$ 
amplitudes (if charge conservation is imposed). Even though this number can be further reduced by 
assuming that the nucleus has good isospin~\cite{Carlson:2015}, the action of pairwise operators 
largely increases the number of components with respect to the initial wave function, thus losing the 
computational advantage of the polynomial scaling with $A$.
However, linear spin/isospin operators do not break the single-particle representation. 
They simply imply rotations of the initial spinors, without generating new amplitudes, as for instance:
\begin{align}
\langle S|\sigma_1^\alpha|\Psi\rangle &=
\sigma_1^\alpha\,\xi_{\alpha_1}(s_1)\,\xi_{\alpha_2}(s_2)\,\xi_{\alpha_3}(s_3)\dots\,\xi_{\alpha_A}(s_A) \nonumber \\
& =\xi_{\alpha_1}(s'_1)\,\xi_{\alpha_2}(s_2)\,\xi_{\alpha_3}(s_3)\dots\,\xi_{\alpha_A}(s_A) \nonumber \\
& =\langle S'|\Psi\rangle .
\end{align}
Quadratic operators can be linearized by using the Hubbard-Stratonovich transformation:
\begin{align}
e^{-\frac{1}{2}\lambda \mathcal O^2}=\frac{1}{\sqrt{2\pi}}\int dx\, e^{-\frac{x^2}{2}+\sqrt{-\lambda}x\mathcal O} ,
\label{eq:hs}
\end{align}
where $x$ are usually called auxiliary fields, and the integral above can
be computed with Monte Carlo techniques, i.e., by sampling points $x$ with  
probability distribution $P(x)=\exp(-x^2/2)$. By using the transformation of~\cref{eq:hs},
Hamiltonians involving up to quadratic operators in spin and isospin can be efficiently
employed in the imaginary-time propagation of a trial wave function of the form of~\cref{eq:wf_s},
retaining the good polynomial scaling with $A$.

\subsection{Propagation of spin/isospin quadratic operators}
\label{sec:p2}
Let us consider the two-body interaction of~\cref{eq:v_ij} up to $p=6$:
\begin{widetext}
\begin{align}
V_{NN}^6&=\sum_{i<j}\Big\{	
 \Big[v_1(r_{ij})+v_2(r_{ij})\,\bm\tau_i\cdot\bm\tau_j\Big]\mathbbm1
+\Big[v_3(r_{ij})+v_4(r_{ij})\,\bm\tau_i\cdot\bm\tau_j\Big]\bm\sigma_i\cdot\bm\sigma_j
+\Big[v_5(r_{ij})+v_6(r_{ij})\,\bm\tau_i\cdot\bm\tau_j\Big]S_{ij}\Big\} , \nonumber \\
&=\sum_{i<j}v_1(r_{ij})+\sum_{i<j}\Big[v_2(r_{ij})\Big]\bm\tau_i\cdot\bm\tau_j
+\sum_{i<j}\sum_{\alpha\beta}\Big[v_3(r_{ij})\,\delta_{\alpha\beta}
+v_5(r_{ij})(3\,\hat{r}_{ij}^\alpha\,\hat{r}_{ij}^\beta-\delta_{\alpha\beta})\Big]\sigma_i^\alpha\sigma_j^\beta \nonumber \\ 
&\quad+\sum_{i<j}\sum_{\alpha\beta}\Big[v_4(r_{ij})\,\delta_{\alpha\beta}
+v_6(r_{ij})(3\,\hat{r}_{ij}^\alpha\,\hat{r}_{ij}^\beta-\delta_{\alpha\beta})\Big]\bm\tau_i\cdot\bm\tau_j\,\sigma_i^\alpha\sigma_j^\beta , \nonumber \\
&=V_{SI}(R)+{1\over2}\sum_{i\ne j}A^{(\tau)}_{ij}\,\bm\tau_i\cdot\bm\tau_j
+{1\over2}\sum_{i\ne j}\sum_{\alpha\beta}A^{(\sigma)}_{i\alpha j\beta}\,\sigma_i^{\alpha}\sigma_j^{\beta}
+{1\over2}\sum_{i\ne j}\sum_{\alpha\beta}A^{(\sigma\tau)}_{i\alpha j\beta} \,\bm\tau_i\cdot\bm\tau_j\,\sigma_i^{\alpha}\sigma_j^{\beta} , \nonumber \\
&=V_{SI}(R)+V_{SD}(R), \label{eq:v_si+sd}
\end{align}
\end{widetext}
where $V_{SI}(V_{SD})$ is the spin/isospin-independent(-dependent) part of the interaction, 
and $A^{(\tau)}_{ij}\;(A\times A)$, $A^{(\sigma)}_{i\alpha j\beta}\;(3A\times 3A)$, 
and $A^{(\sigma\tau)}_{i\alpha j\beta}\;(3A\times 3A)$
are real and symmetric matrices.
As such, these matrices can be diagonalized:
\begin{align}
\sum_{j}A^{(\tau)}_{ij}\,\psi_{n,j}^{(\tau)}&=\lambda_n^{(\tau)}\,\psi_{n,i}^{(\tau)} , \nonumber \\
\sum_{j\beta}A^{(\sigma)}_{i\alpha j\beta}\,\psi_{n,j\beta}^{(\sigma)}&=\lambda_n^{(\sigma)}\,\psi_{n,i\alpha}^{(\sigma)} , \nonumber \\
\sum_{j\beta}A^{(\sigma\tau)}_{i\alpha j\beta}\,\psi_{n,j\beta}^{(\sigma\tau)}&=\lambda_n^{(\sigma\tau)}\,\psi_{n,i\alpha}^{(\sigma\tau)} ,
\end{align}
and it is possible to define a new set of operators expressed in terms of their eigenvectors: 
\begin{align}
\mathcal O_{n\alpha}^{(\tau)}&=\sum_{j}\tau_j^\alpha\,\psi_{n,j}^{(\tau)} , \nonumber \\
\mathcal O_{n}^{(\sigma)}&=\sum_{j\beta}\sigma_j^\beta\,\psi_{n,j\beta}^{(\sigma)} , \nonumber \\
\mathcal O_{n\alpha}^{(\sigma\tau)}&=\sum_{j\beta}\tau_j^\alpha\sigma_j^\beta\,\psi_{n,j\beta}^{(\sigma\tau)} ,  
\end{align}
such that the spin/isospin-dependent part of~\cref{eq:v_si+sd} can be recast as:
\begin{align}
V_{SD}(R)&= 
 {1\over2}\sum_{\alpha=1}^3\sum_{n=1}^{A} \lambda_n^{(\tau)}\Big(\mathcal O_{n\alpha}^{(\tau)}\Big)^2
+{1\over2}\sum_{n=1}^{3A} \lambda_n^{(\sigma)}\Big(\mathcal O_n^{(\sigma)}\Big)^2 \nonumber \\
&\quad+{1\over2}\sum_{\alpha=1}^3\sum_{n=1}^{3A} \lambda_n^{(\sigma\tau)}\Big(\mathcal O_{n\alpha}^{(\sigma\tau)}\Big)^2 .
\end{align}

The potential written in this form contains only quadratic operators in spin/isospin. 
We can thus use the Hubbard-Stratonovich transformation of~\cref{eq:hs} to write the
propagator of the $V_{NN}^6$ interaction acting on a configuration $|RS\rangle$ as:
\begin{widetext}
\begin{align}
e^{-V_{NN}^6\delta\tau}|RS\rangle=e^{-V_{SI}(R)\delta\tau}\prod_{m=1}^{15A}\frac{1}{\sqrt{2\pi}}
\displaystyle\int dx_m\,e^{\frac{x_m^2}{2}}\,e^{\sqrt{-\lambda_m\delta\tau}\,x_m \mathcal O_m}|RS\rangle
=|RS'\rangle ,
\end{align}
\end{widetext}
where 15 auxiliary fields are needed for each nucleon, 3 for $\tau$ operators, 3 for $\sigma$,
and 9 for $\sigma\tau$. The propagation (rotation) of spinors depends upon the sampling
of the auxiliary fields $X=\{x_m\}$, so does the new spin/isospin configurations $S'\equiv S'(X)$.
The full short-time propagator, which includes both kinetic and potential energies, can finally be expressed as:
\begin{widetext}
\begin{align}
G(R',R,S'(X),S,\delta\tau)=\langle R'S'|\Big({m\over2\pi\hbar^2\delta\tau}\Big)^{3A\over2}e^{-{m(R-R')^2\over2\hbar^2\delta\tau}}
e^{-(V_{SI}(R)-E_T)\delta\tau}
\prod_{m=1}^{15A}{1\over\sqrt{2\pi}}\int dx_m\,e^{-{x_m^2\over2}}
e^{\sqrt{-\lambda_m\delta\tau}\,x_m \mathcal O_m}|RS\rangle.
\label{eq:prop}
\end{align}
\end{widetext}

Note that the above expressions refer to the simple propagator $\exp[-T\delta\tau]\exp[-(V-E_T)\delta\tau]$.
In practice, we sample the more accurate propagator 
$\exp[-(V-E_T)\delta\tau/2]\exp[-T\delta\tau]\exp[-(V-E_T)\delta\tau/2]$, which implies
two sets of rotations in $\delta\tau/2$: the first depending on $R$, and the second on 
the diffused $R'$, for a total of 30 auxiliary fields.
Compared to the GFMC method, where the coordinates are sampled and the spin and isospin states are 
explicitly included and summed, in AFDMC, spin and isospin are also sampled via 
Hubbard-Stratonovich rotations. This largely reduces the computational cost of the
imaginary-time propagation of a many-body wave function, allowing one to calculate nuclei
more efficiently up \isotope[12]{C}, and to go beyond $A=12$.

\subsection{Propagation of spin-orbit operators}
\label{sec:pls}
The spin-orbit operator reads
\begin{align}
v_{LS}(r_{ij})=v_7(r_{ij})\,\vb{L}\cdot\vb{S} ,
\end{align}
where $\vb{L}$ and $\vb{S}$ are defined in~\cref{eq:L,eq:S}, respectively.
As shown in Ref.~\cite{Pieper:1998}, one way to evaluate
the propagator for spin-orbit operators is to consider the expansion at first order in $\delta\tau$
\begin{align}
	e^{-v_7(r_{ij})\,\vb{L}\cdot\vb{S}\,\delta\tau}\approx\mathbbm1-v_7(r_{ij})\,\vb{L}\cdot\vb{S}\,\delta\tau ,
\end{align}
acting on the free propagator $G_0$ of~\cref{eq:g0}.
The resulting propagator is
\begin{align}
G_{LS}\approx e^{\sum_{i\ne j}\frac{1}{8i}\frac{2m}{\hbar^2}v_7(r_{ij})(\vb{r}_i-\vb{r}_j)
\times(\bm\Delta\vb{r}_i-\bm\Delta\vb{r}_j)\cdot(\bm\sigma_i+\bm\sigma_j)} ,
\end{align}
where $\bm\Delta\vb{r}_i=\vb{r}_i-\vb{r}_i'$ is the difference of the particle position before and after the
action of the free propagator $G_0$.
Note that the above propagator is only linear in the spin, i.e., it does not require any
auxiliary field to be sampled.
However, it can be shown that it induces spurious counter terms~\cite{Sarsa:2003}.
These can be removed by using the modified propagator:
\begin{align}
G_{LS}\approx &\,e^{\sum_{i\neq j}\frac{1}{4i}\frac{m}{\hbar^2\delta\tau}v_7(r_{ij})
[\vb{r}_{ij}\times\bm\Delta\vb{r}_{ij}]\cdot\bm\sigma_i}
\nonumber \\
&\times e^{-\frac{1}{2}\left[\sum_{i\neq j}\frac{1}{4i}\frac{m}{\hbar^2}v_7(r_{ij})
[\vb{r}_{ij}\times\bm\Delta\vb{r}_{ij}]\cdot\bm\sigma_i\right]^2} .
\end{align}
This alternative version of the spin-orbit propagator contains quadratic spin operators,
and thus it requires additional Hubbard-Stratonovich fields to be sampled, but it is 
correct at order $\delta\tau$.

\subsection{Propagation of three-body forces}
\label{sec:p3}
Several terms of the $3N$ interaction (\cref{eq:v_ijk}) can be directly included in the AFDMC propagator. 
These are $V_a^{2\pi,P}$, $V^{2\pi,S}$, $V_D$, and $V_E$ of~\cref{eq:anti,eq:tm,eq:vd,eq:ve},
which correspond to terms involving only quadratic spin and isospin operators.
These have the same operator structure as the spin/isospin-dependent part of the
two-body potential (\cref{eq:v_si+sd}). The dependence on the third particle $k$ enters only in the radial
functions ${\cal X}_{i\alpha j\beta}$, ${\cal Z}_{ij\alpha}$, and $\delta_{R_0}(r)$,
which can be absorbed in the definition of the
matrices $A^{(\tau)}_{ij}$ and $A^{(\sigma\tau)}_{i\alpha j\beta}$.

The structure of $V_c^{2\pi,P}$ contains instead cubic spin and isospin operators, 
and the Hubbard-Stratonovich transformation of~\cref{eq:hs} cannot be applied.
It follows that these terms cannot be exactly included in the standard AFDMC propagation.
It may be possible to invoke more complicated algorithms to sample them, 
but the imaginary-time step will need to be higher order in $\delta\tau$.
However, their expectation value can always be calculated, and it can be used to derive
an approximate three-body propagator for $V_c^{2\pi,P}$.

Let us define an effective Hamiltonian $H'$ that can be exactly included in 
the AFDMC propagation:
\begin{align}
H'=H-V_c^{2\pi,P}+\alpha_1 V_a^{XX}+\alpha_2 V_D^{X\delta}+\alpha_3 V_E .
\label{eq:h'}
\end{align}
The three constants $\alpha_i$ are adjusted in order to have:
\begin{align}
\langle V_c^{XX}\rangle & \approx\langle\alpha_1 V_a^{XX}\rangle , \nonumber \\
\langle V_c^{X\delta}\rangle & \approx\langle\alpha_2 V_D^{X\delta}\rangle , \nonumber  \\
\langle V_c^{\delta\delta}\rangle & \approx\langle\alpha_3 V_E\rangle , 
\label{eq:pert}
\end{align}
where $\langle\,\cdots\rangle$ indicates the average over the wave function
(see~\cref{sec:obs}), and the identifications are suggested by the similar 
ranges and functional forms.

Once the ground state $\Psi_0'$ of $H'$ is calculated via the AFDMC
imaginary-time propagation, the expectation value of the Hamiltonian $H$ is given by
\begin{align}
\!\!\langle H\rangle &\approx\langle\Psi_0'|H'|\Psi_0'\rangle+\langle\Psi_0'|H-H'|\Psi_0'\rangle \nonumber \\
& \approx\langle H'\rangle+\langle V_c^{2\pi,P}\!-\alpha_1V_a^{XX}\!-\alpha_2V_D^{X\delta}\!-\alpha_3V_E\rangle \nonumber \\
& \approx\langle H'\rangle+\langle V_{\rm pert}\rangle ,
\label{eq:<h3b>}
\end{align}
where the last term is evaluated perturbatively, meaning that its expectation
value is calculated, even though not all the operators are included in the propagator $(V_c^{2\pi,P})$.
By opportunely adjusting the constants $\alpha_i$ of~\cref{eq:pert}, we ensure that
the correction $\langle V_{\rm pert}\rangle$ is small compared to $\langle H'\rangle$.
A similar approach is used in the GFMC method to calculate the small nonlocal terms that are 
present in the AV18 interaction. In that case the difference $v_8'-v_{18}$
is calculated as a perturbation~\cite{Pudliner:1997}.

\subsection{Importance sampling}
\label{sec:is}
Diffusion Monte Carlo algorithms, such as the GFMC and AFDMC methods, are much more efficient when importance sampling
techniques are also implemented.
In fact, sampling spatial and spin/isospin configurations according to $G(R',R,S'(X),S,\delta\tau)$ 
might not always be efficient. For instance, consider the case of a strongly repulsive interaction 
at short distances. In such a situation, sampling the spatial coordinates according to the kinetic 
energy only is not an optimal choice because no information about the interaction is included in 
sampling the paths, but only through the weights associated with the configurations. 
As a result, an inefficiently sampled path might have a very small weight, making its contribution 
very small along the imaginary time.

Suppose that we construct a positive definite wave function $\Psi_G$ close to that of the true ground 
state of the Hamiltonian $H$. $\Psi_G$ can be used to guide the imaginary-time evolution by defining a
better propagator compared to that of~\cref{eq:imtimeprop}, to be used to sample coordinates and 
spin/isospin configurations:
\begin{widetext}
\begin{align}
\langle\Psi_G|R'S'\rangle\langle R'S'|\Psi(\delta\tau)\rangle
&=\displaystyle\int dR\,G(R',R,S'(X),S,\delta\tau)\,\langle\Psi_G|R'S'(X)\rangle\langle RS|\Psi_T(0)\rangle
\nonumber \\
&=\displaystyle\int dR\,G(R',R,S'(X),S,\delta\tau)\frac{\langle\Psi_G|R'S'(X)\rangle}{\langle\Psi_G|RS\rangle}
\langle\Psi_G|RS\rangle\langle RS|\Psi_T(0)\rangle .
\end{align}
\end{widetext}
Note that if $\Psi_G$ is positive definite, the above propagation does not change the 
variance of the computed observables.

In typical DMC calculations the modified propagator is sampled by shifting the Gaussian in
the free propagator, and then including the local energy in the weight of the configuration
(see, e.g., Ref.~\cite{Foulkes:2001}). A similar approach has also been used in AFDMC calculations in the past. 
However, in the latest implementation of the AFDMC method, a much more efficient way to implement the
importance sampling propagator is used.

The goal is to sample the modified propagator:
\begin{align}
G(R',R,S'(X),S,\delta\tau)\frac{\langle\Psi_G|R'S'(X)\rangle}{\langle\Psi_G|RS\rangle} .
\end{align}
We first sample a set of coordinate displacements $\Delta R$ according
to~\cref{eq:prop} and a
set of auxiliary fields $X$ from Gaussian distributions.
Since the propagator $G$ implies the Gaussian sampling for the kinetic energy and 
for the auxiliary fields, sampling $\Delta R$ and $X$ has the same probability
of sampling $-\Delta R$ and $-X$.
Driven by this observation, we calculate the ratios:
\begin{align}
w_1&=\frac{\langle\Psi_G|R+\Delta R,S'(X)\rangle}{\langle\Psi_G|RS\rangle}e^{-[V_{SI}(R+\Delta R)-E_T]\delta\tau},
\nonumber \\
w_2&=\frac{\langle\Psi_G|R-\Delta R,S'(X)\rangle}{\langle\Psi_G|RS\rangle}e^{-[V_{SI}(R-\Delta R)-E_T]\delta\tau},
\nonumber \\
w_3&=\frac{\langle\Psi_G|R+\Delta R,S'(-X)\rangle}{\langle\Psi_G|RS\rangle}e^{-[V_{SI}(R+\Delta R)-E_T]\delta\tau},
\nonumber \\
w_4&=\frac{\langle\Psi_G|R-\Delta R,S'(-X)\rangle}{\langle\Psi_G|RS\rangle}e^{-[V_{SI}(R-\Delta R)-E_T]\delta\tau},
\label{eq:w_i}
\end{align}
where $V_{SI}$ is the spin/isospin-independent part of the interaction.
We then sample one of the above choices according to the ratios $w_i$.
Finally, the total weight of the new configuration is given by
\begin{align}
W=\frac{1}{4}\sum_i w_i ,
\end{align}
and $W$ is used for branching as in the standard DMC method~\cite{Carlson:2015}.

\subsection{Observables}
\label{sec:obs}
The expectation value of an observable $\mathcal O$ is calculated by using the sampled configurations $R_iS_i$ as:
\begin{align}
\displaystyle\langle \mathcal O(\tau)\rangle=\frac{\displaystyle\sum_i \frac{\langle R_iS_i|\mathcal O|\Psi_T\rangle}{W}\frac{W}{\langle R_iS_i|\Psi_T\rangle}}
{\displaystyle\sum_i\frac{W}{\langle R_iS_i|\Psi_T\rangle}}.
\label{eq:obs}
\end{align}
The above expression is valid only for observables that commute with the Hamiltonian.
For other observables, such as radii and densities, expectation values are often calculated from mixed estimates 
\begin{align}
\langle\mathcal O(\tau)\rangle\approx2\frac{\langle\Psi_T|\mathcal O|\Psi(\tau)\rangle}{\langle\Psi_T|\Psi(\tau)\rangle}
-\frac{\langle\Psi_T|\mathcal O|\Psi_T\rangle}{\langle\Psi_T|\Psi_T\rangle}	,
\label{eq:mix}
\end{align}
where the first term corresponds to the DMC expectation value, and the second term is the VMC one.
\Cref{eq:mix} is valid for diagonal matrix elements, but it can be generalized to the case of off-diagonal matrix 
elements, e.g., in transition matrix elements between different initial and final states
(see Ref.~\cite{Pervin:2007}). 

Note that the extrapolation above is small for accurate wave functions.
This is the case, for instance, for closed-shell nuclei and single operators.
For open-shell systems, particularly for halo nuclei, the information encoded in the trial
wave function may not be as accurate as that for simpler systems. This can result in
a nonnegligible extrapolation of the mixed expectation value.
An example of this behavior is provided by the nuclear radius, 
the VMC expectation value of which is typically larger than the DMC one
for open-shell systems. One way to reduce the extrapolation of the mixed estimate for the radius
is to use a penalty function during the optimization of the variational parameters 
in the trial wave function. This penalty function sets a constraint on the VMC radius so as to
adjust its expectation value close to the DMC estimate, thus reducing the extrapolation.

\subsection{Constrained and unconstrained evolution}
\label{sec:cp}
The fact that the weight $W$ is always real and positive and that $\Psi_T$ is complex 
makes the denominator of~\cref{eq:obs} average quickly to zero. 
This is the well known sign problem in DMC methods.
One way to avoid the sign problem is to use a constraint during the imaginary-time evolution. 
In practice, a configuration is given zero weight (thus it is dropped during branching) 
if its real part changes sign.

In our implementation of the AFDMC method, we follow Ref.~\cite{Zhang:2003}. 
In sampling the propagator, we calculate the weights $w_i$ of~\cref{eq:w_i} as
\begin{align}
\frac{\langle\Psi_G|(R',S'(X)\rangle}{\langle\Psi_G|RS\rangle}\rightarrow
\Re\left\{\frac{\langle\Psi_T|(R',S'(X)\rangle}{\langle\Psi_T|RS\rangle}\right\},
\end{align}
and we then apply the constraint by assigning zero weight to a move
that results in a negative ratio. This is analogous to the constrained-path 
approximation~\cite{Zhang:1997}, but for complex wave functions and propagators.

This constrained evolution does not suffer a sign problem, but it makes the final result
dependent on the choice of $\Psi_T$. Moreover, it implies that the calculated energy is not necessarily
an upperbound to the true ground-state energy, as is the case of the fixed-node approximation 
in real space~\cite{Ortiz:1993,Foulkes:2001}.

The results given by the constrained evolution can be improved by releasing the 
constraint and following the unconstrained evolution.
After a set of configurations is generated using the constraint, the guiding function
is taken as
\begin{align}
\langle\Psi_G|RS\rangle=\Re\big\{\langle\Psi_G|RS\rangle\big\}+\alpha\Im\big\{\langle\Psi_G|RS\rangle\big\},
\end{align}
where $\alpha$ is a small arbitrary constant.
This ensures that the ratio in the weights $w_i$ of~\cref{eq:w_i} is always positive and real.
The propagation continues then according to the modified $\langle\Psi_G|RS\rangle$, and 
observables are calculated as before according to~\cref{eq:obs}.
In several cases the expectation value $\langle O\rangle$ reaches a stable value independent
of imaginary time before the signal-to-noise ratio goes to zero, and the result is exact 
within the statistical uncertainty.
This is the case for light systems, $A\le4$. For larger nuclei the variance grows much faster 
as a function of the imaginary time, so that the unconstrained evolution cannot always be 
followed until $\langle O\rangle$ reaches a plateau. 
In these cases, the final result is
extrapolated using an exponential fit as in Ref.~\cite{Pudliner:1997}. We found that a 
single-exponential form with free-sign coefficients yields the most stable fits in our case. 
Such a form has been used to obtain all the quoted results.
Examples of unconstrained evolution are provided in~\cref{sec:constr}.

\section{Trial wave function}
\label{sec:wf}
The AFDMC trial wave function we use takes the form:
\begin{widetext}
\begin{align}
\langle RS|\Psi\rangle=\langle RS|\prod_{i<j}f^1_{ij}\,\prod_{i<j<k}f^{3c}_{ijk}\,
\left[\mathbbm1+\sum_{i<j}\sum_{p=2}^6 f^p_{ij}\,\mathcal O_{ij}^p\, f_{ij}^{3p}
+\sum_{i<j<k}U_{ijk}\right]|\Phi\rangle_{J^\pi,T} ,
\label{eq:psi}
\end{align}
\end{widetext}
where $|RS\rangle$ represents the sampled $3A$ spatial coordinates and the $4A$ spin/isospin amplitudes
for each nucleon, and the pair correlation functions $f^{p=1,6}_{ij}\equiv f^{p=1,6}(r_{ij})$ are obtained as the solution
of Schr\"odinger-like equations in the relative distance between two particles, as explained in Ref.~\cite{Carlson:2015}. 
The two spin/isospin-independent functions $f^{3c}_{ijk}$ and $f^{3p}_{ij}$ are defined as
\begin{align}
f^{3c}_{ijk}&=1+q_1^c\,\vb{r}_{ij}\cdot\vb{r}_{ik}\,\vb{r}_{ji}\cdot\vb{r}_{jk}\,
\vb{r}_{ki}\cdot\vb{r}_{kj}\,e^{-q_2^c(r_{ij}+r_{ik}+r_{jk})} ,
\nonumber \\
f_{ij}^{3p}&=\prod_k\left[1-q_1^p(1-\vb{r}_{ik}\cdot\vb{r}_{jk})\,e^{-q_2^p(r_{ij}+r_{ik}+r_{jk})}\right] ,
\end{align}
and they are introduced to reduce the strength of the spin/isospin-dependent pair correlation functions when 
other particles are nearby~\cite{Pudliner:1997}.
Finally, three-body spin/isospin-dependent correlations are also included as 
\begin{align}
U_{ijk}=\sum_n \epsilon_n V_{ijk}^n(\alpha_n r_{ij},\alpha_n r_{ik},\alpha_n r_{jk}) ,
\end{align}
where the terms $V^n_{ijk}$ are the same as the $3N$ interactions of~\cref{eq:v_ijk}, 
$\epsilon_n$ are potential quenching factors, and $\alpha_n$ are coordinate scaling factors.
In the correlations above, we include the four terms $V_a^{2\pi,P}$, $V^{2\pi,S}$,
$V_D$, and $V_E$. $V_c^{2\pi,P}$ can also be implemented in the trial wave function,
but since its structure involves three-body spin/isospin operators, its inclusion 
results in a severely larger computational cost.

The term $|\Phi\rangle$ is taken as a shell-model-like wave function. 
It consists of a sum of Slater determinants constructed using single-particle orbitals:
\begin{align} 
\langle RS|\Phi\rangle_{J^\pi,T} = \sum_n c_n\Big[\sum \mathcal C_{J\!M}\,\mathcal D\big\{\phi_\alpha(\vb{r}_i,s_i)\big\}_{J,M}\Big]_{J^\pi,T} ,
\end{align}
where $\vb{r}_i$ are the spatial coordinates of the nucleons, and $s_i$ represents their spinor.
$J$ is the total angular momentum, $M$ its projection, $T$ the total isospin, and $\pi$ the parity.
The determinants $\mathcal D$ are coupled with Clebsch-Gordan coefficients $\mathcal C_{J\!M}$ 
in order to reproduce the experimental total angular momentum, total isospin, and parity $(J^\pi,T)$. 
The $c_n$ are variational parameters multiplying different components having the same quantum numbers. 
Each single-particle orbital $\phi_\alpha$ consists of a radial function multiplied by the spin/isospin trial states:
\begin{align}
\phi_\alpha(\vb{r}_i,s_i)=\Phi_{nj}(r_i)\left[Y_{l,m_l}(\hat{\vb{r}}_i)\chi_\gamma(s_i)\right]_{j,m_j} ,
\label{eq:phi}
\end{align}
where the spherical harmonics $Y_{l,m_l}(\hat{\vb{r}}_i)$ are coupled to the spin state $\chi_\gamma(s_i)$
in order to have single-particle orbitals in the $j$ basis.
The radial parts $\Phi(r)$ are obtained from the bound-state solutions of the Woods-Saxon 
wine-bottle potential:
\begin{align}
v(r)=V_s\left[\frac{1}{1+e^{(r-r_s)/a_s}}+\alpha_s\,e^{-(r/{\rho_s})^2}\right] ,
\end{align}
where the five parameters $V_s$, $r_s$, $a_s$, $\alpha_s$, and $\rho_s$ can be different for orbitals
belonging to different states, such as $1S_{1/2}$, $1P_{3/2}$, $1P_{1/2}$,\ldots, and they are 
optimized in order to minimize the variational energy.
Finally, the spin/isospin trial states are represented in the $|p\uparrow\rangle$, $|p\downarrow\rangle$,
$|n\uparrow\rangle$, $|n\downarrow\rangle$ basis $(|\chi_{\gamma=1,4}\rangle)$. 
The spinors are specified as:
\begin{align}
|s_i\rangle \equiv \left(\begin{array}{c} 
a_i \\ b_i \\ c_i \\ d_i
\end{array}\right)
=a_i|p\uparrow\rangle+b_i|p\downarrow\rangle+c_i|n\uparrow\rangle+d_i|n\downarrow\rangle ,
\end{align}
and the trial spin/isospin states are taken to be:
\begin{align}
\chi_1(s_i)&=\langle s_i|\chi_1\rangle=\langle s_i|(1,0,0,0)\rangle=a_i,\nonumber \\
\chi_2(s_i)&=\langle s_i|\chi_2\rangle=\langle s_i|(0,1,0,0)\rangle=b_i,\nonumber \\
\chi_3(s_i)&=\langle s_i|\chi_3\rangle=\langle s_i|(0,0,1,0)\rangle=c_i,\nonumber \\
\chi_4(s_i)&=\langle s_i|\chi_4\rangle=\langle s_i|(0,0,0,1)\rangle=d_i .
\end{align}

Let us consider a system with $K$ states. According to the definitions above, 
a single Slater determinant $\mathcal D\equiv\mathcal D\big\{\phi_\alpha(\vb{r}_i,s_i)\big\}_{J,M}$ is constructed as:
\begin{align}
\mathcal D=
\left|
\begin{array}{cccc}
a_1\phi_1(\vb{r}_1) & a_2\phi_1(\vb{r}_2) & \dots & a_A\phi_1(\vb{r}_A) \\
a_1\phi_2(\vb{r}_1) & a_2\phi_2(\vb{r}_2) & \dots & a_A\phi_2(\vb{r}_A) \\
\dots & \dots & \dots & \dots \\
b_1\phi_1(\vb{r}_1) & b_2\phi_1(\vb{r}_2) & \dots & b_A\phi_1(\vb{r}_A) \\
b_1\phi_2(\vb{r}_1) & b_2\phi_2(\vb{r}_2) & \dots & b_A\phi_2(\vb{r}_A) \\
\dots & \dots & \dots & \dots \\
d_1\phi_1(\vb{r}_1) & d_2\phi_1(\vb{r}_2) & \dots & d_A\phi_1(\vb{r}_A) \\
\dots & \dots & \dots & \dots \\
d_1\phi_K(\vb{r}_1) & d_2\phi_K(\vb{r}_2) & \dots & d_A\phi_K(\vb{r}_A) \\
\end{array}
\right| . 
\end{align}
For \isotope[16]{O}\,$(0^+,0)$, for instance, the number of states is four: 
one $1S_{1/2}$, two $1P_{3/2}$, and one $1P_{1/2}$. Each of them
can accommodate two spins and two isospin states, and the full 
$\langle RS|\Phi\rangle_{0^+,0}$
wave function can be written as a single Slater determinant. 
For open-shell systems instead, many Slater determinants need to be 
included in order to have a good trial wave function with the proper $(J^\pi,T)$.
For $A=6$ systems, e.g., including single-particle orbitals up to the 
$sd$-shell there are ten possible states: 
one $1S_{1/2}$, two $1P_{3/2}$, one $1P_{1/2}$, three $1D_{5/2}$, 
two $1D_{3/2}$, and one $2S_{1/2}$. These can be combined in 9
different Slater determinants in order to have the \isotope[6]{He}\,$(0^+,1)$ wave function, 
or in 32 Slater determinants to make \isotope[6]{Li}\,$(1^+,0)$. 
Finally, for \isotope[12]{C}\,$(0^+,0)$, by considering only $K=4$ as for 
\isotope[16]{O}\,$(0^+,0)$, the number of Slater determinants needed to build a $(0^+,0)$ 
wave function is already 119, making it computationally challenging to include $sd$-shell orbitals for $A=12$.

The trial wave function of~\cref{eq:psi} contains a sum over pair correlation functions, 
meaning that only one pair of nucleons $ij$ is correlated at a time (linear correlations).
This is different from the GFMC wave function~\cite{Carlson:2015}, 
where all pairs are correlated at the same time. 
In the AFDMC method, this same construction would, however, forbid the application of the 
Hubbard-Stratonovich transformation, justifying the choice of~\cref{eq:psi}. 
An improved AFDMC two-body wave function could include linear and
quadratic pair correlations: 
\begin{widetext}
\begin{align}
\langle RS|\Psi\rangle_{2b}=\langle RS|\prod_{i<j}f^1_{ij}\,
\left[\mathbbm1+\sum_{i<j}\sum_{p=2}^6 f^p_{ij}\,\mathcal O_{ij}^p\,
+\sum_{i<j}\sum_{p=2}^6 f^p_{ij}\,\mathcal O_{ij}^p \sum_{\substack{k<l\\ij\ne kl}}\sum_{q=2}^6 f^q_{kl}\,\mathcal O_{kl}^q\,\right]|\Phi\rangle_{J^\pi,T} ,
\label{eq:psi2}
\end{align}
\end{widetext}
where the sum over $kl$ includes all nucleon pairs except when $k=i$ and $l=j$. 
The $f^{p,q}_{ij}$ functions are solved for as before, and the operators 
$\mathcal{O}^{p,q}_{ij}$ are the same as in~\cref{eq:psi}. 
Although the two-body wave function of~\cref{eq:psi2} contains all quadratic correlations, 
most of the relevant physics is captured with a subset of these correlations, corresponding
to the action of the $\mathcal O_{ij}^{p,q}$ operators on four distinct 
particles (so-called independent pair correlations). Since these correlations never act on the same 
particle, all the $\mathcal{O}^{p,q}_{ij}$ operators commute, removing the need for an explicit 
symmetrization of the wave function. Such a wave function could, in principle, improve
the energy expectation value for large systems, but the computational cost of its evaluation is 
significantly higher than for a wave function with linear correlations only.
In fact, the cost of computing expectation values of two-body operators on
a two-body wave function of the form of~\cref{eq:psi2} is proportional to $A^4$ for linear correlations, 
and to $A^6$ for quadratic correlations.
For this reason in the present work we consider only linear two-body correlations in the wave function, 
and we present a test study of quadratic correlations in~\cref{sec:psi2}.

\section{Results}
\label{sec:res}

\subsection{Test of constrained and unconstrained evolution}
\label{sec:constr}
\begin{figure}[b]
\includegraphics[width=\linewidth]{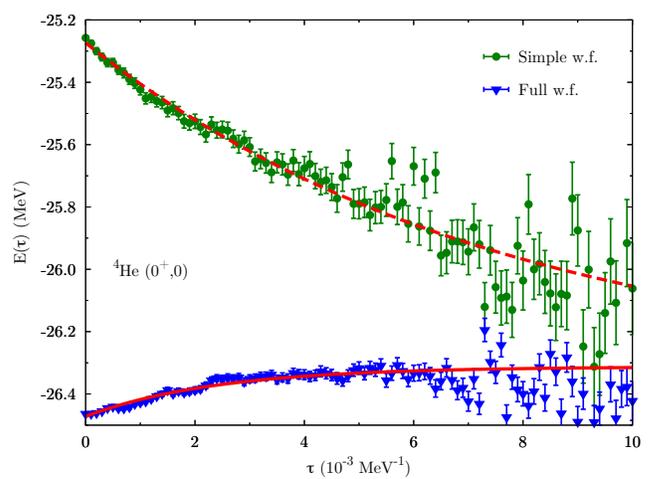}
\caption[]{Energy of \isotope[4]{He} as a function of 
imaginary time after releasing the constraint for the AV6$'$ potential.
The two data sets refer to the two different wave functions of~\cref{tab:tr}.
Red lines are exponential fits to the Monte Carlo results.}
\label{fig:tr_he4}
\end{figure}

As introduced in \cref{sec:cp}, the energy (and other observables) calculated with the
AFDMC method during the constrained evolution is dependent on the choice of $\Psi_T$.
This is shown in~\cref{tab:tr} where the energy of \isotope[4]{He} 
is calculated for the Argonne $v_6'$ (AV6$'$) potential~\cite{Wiringa:2002} 
employing different trial wave functions. 
Full w.f. refers to the wave function of~\cref{eq:psi} where all the two-body 
correlations are included.
Simple w.f. is instead a simplified wave function where only the central and
$p=2,5$ operator correlations are used, the strength of the latter 
$(\mathcal O_{ij}^5=S_{ij}\,\bm\tau_i\cdot\bm\tau_j)$ being artificially reduced by 
a factor 3 after the optimization process. At the variational level it is evident how
the simplified wave function is not the optimal choice for $\Psi_T$, as the energy
expectation value is much higher than for the fully optimized wave function. For both choices
of $\Psi_T$, the constrained evolution reduces the binding energy, moving towards the
GFMC reference value for the same potential (see~\cref{tab:av6c}), but the results are still inconsistent.
It is only the unconstrained evolution that brings the results for both wave functions 
in agreement within statistical errors. This is also shown in~\cref{fig:tr_he4}, where the 
AFDMC energy is plotted as a function of imaginary time for the unconstrained evolution.

\setlength{\tabcolsep}{8pt}
\begin{table}[tb]
\centering
\caption[]{\isotope[4]{He} ground-state energies for the AV6$'$ potential and different 
trial wave functions (see text for details). 
C(U) refers to the constrained(unconstrained) evolution.
Errors are statistical. Results are in MeV.}
\begin{tabular}{lcc}
\hline\hline
Energy & Simple w.f. & Full w.f. \\
\hline
$E_{\rm VMC}$           & $-9.49(5)$   & $-23.35(1)$ \\
$E_{\rm AFDMC}^{\rm C}$ & $-25.28(3)$  & $-26.45(1)$ \\
$E_{\rm AFDMC}^{\rm U}$ & $-26.34(12)$ & $-26.31(4)$ \\
\hline\hline
\end{tabular}
\label{tab:tr}
\end{table}
\setlength{\tabcolsep}{8pt}

We report in~\cref{tab:av6c} the constrained and unconstrained energies for $A=3,4,6$ 
employing the AV6$'$ potential, in comparison with the GFMC results for the same
interaction~\cite{Wiringa:2002}. It is interesting to note that constrained energies do not always satisfy 
the variational principle, as anticipated in~\cref{sec:cp}. This is seen, e.g., in \isotope[3]{H}
and \isotope[4]{He}, for which the constrained energy is below the GFMC prediction, considered to be the 
exact solution for the given potential. However, once the unconstrained evolution is performed, the
AFDMC and GFMC results agree within $1\%$ or less.

\begin{table}[tb]
\centering
\caption[]{Ground state energies for $A=3,4,6$ employing the AV6$'$ potential. 
Errors are statistical. Results are in MeV.}
\begin{tabular}{lccc}
\hline\hline
$\isotope[A]{Z}\,(J^\pi,T)$ & $E^{\rm C}_{\rm AFDMC}$ & $E^{\rm U}_{\rm AFDMC}$ & $E_{\rm GFMC}$ \\
\hline
\isotope[3]{H}\,$(\frac{1}{2}^+,\frac{1}{2})$ & $-8.08(1)$  & $-7.95(2)$   & $-7.95(2)$  \\
\isotope[4]{He}\,$(0^+,0)$                    & $-26.45(1)$ & $-26.31(4)$  & $-26.15(2)$ \\
\isotope[6]{Li}\,$(1^+,0)$                    & $-28.09(4)$ & $-28.26(10)$ & $-28.37(4)$ \\
\hline\hline
\end{tabular}
\label{tab:av6c}
\end{table}

\begin{figure}[htb]
\includegraphics[width=\linewidth]{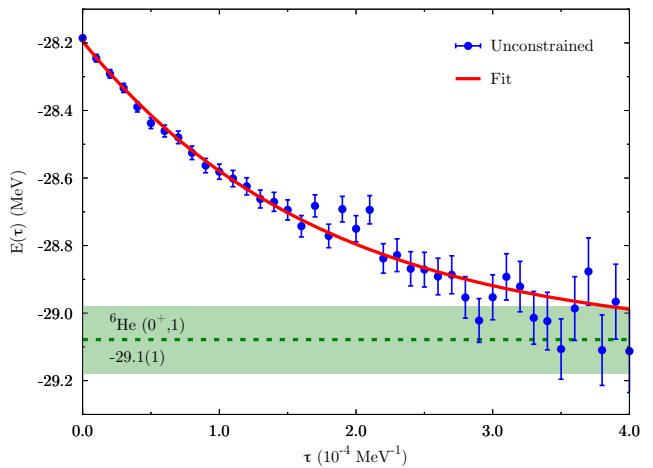}
\caption[]{\isotope[6]{He} unconstrained evolution for the local chiral potential at
N$^2$LO $(E\tau)$ with cutoff $R_0=1.2\,\rm fm$. Data points refer to the expectation 
value of $H'$, \cref{eq:h'}.}
\label{fig:tr_he6}
\end{figure}

\begin{figure}[htb]
\includegraphics[width=\linewidth]{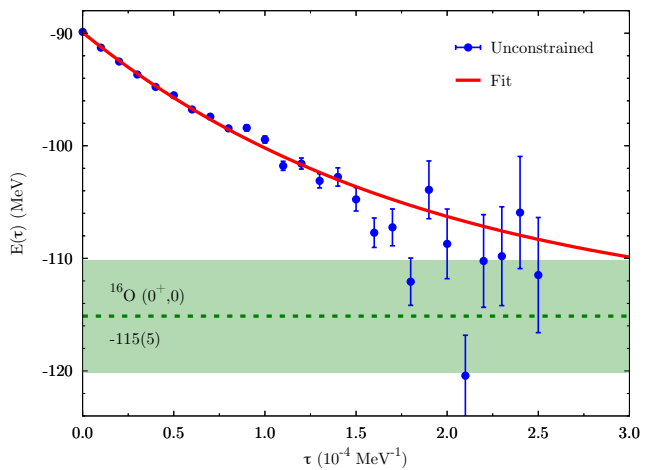}
\caption[]{\isotope[16]{O} unconstrained evolution for the local chiral potential at
N$^2$LO $(E\tau)$ with cutoff $R_0=1.0\,\rm fm$. Data points refer to the expectation 
value of $H'$, \cref{eq:h'}.}
\label{fig:tr_o16}
\end{figure}

In~\cref{fig:tr_he6,fig:tr_o16} we show two examples of unconstrained calculation for larger systems,
\isotope[6]{He} and \isotope[16]{O} respectively, employing realistic two- plus three-body interactions. 
We use the local chiral potential at N$^2$LO with cutoff $R_0=1.2\,\rm fm$ for 
\isotope[6]{He} and $R_0=1.0\,\rm fm$ for \isotope[16]{O}. The employed wave functions include 
all two- and three-body correlations, and for \isotope[6]{He} we include single-particle orbitals 
up to the $sd$-shell. In general, the larger the system, the shorter 
the imaginary-time evolution that can be followed before the variance becomes too large. 
This is particularly evident in \isotope[16]{O}, for which the unconstrained evolution 
can be satisfactorily performed up to $2.5\times10^{-4}\,\rm MeV^{-1}$, 
compared to the $4\times10^{-4}\,\rm MeV^{-1}$
for \isotope[6]{He} of~\cref{fig:tr_he6} with the same interaction, and to the 
$5\times10^{-3}\,\rm MeV^{-1}(10^{-2}\,\rm MeV^{-1})$ 
for \isotope[4]{He} (with AV6$'$ of~\cref{fig:tr_he4}).
For example, at $\tau=2\times10^{-4}\,\rm MeV^{-1}$ the statistical error per nucleon is  
$0.01\,\rm MeV$ for \isotope[4]{He} and \isotope[6]{Li}, and $0.19\,\rm MeV$ for \isotope[16]{O}.
This is a direct consequence of the quality of the employed wave function. 
For small nuclei, the wave function of~\cref{eq:psi} provides a good description of the
system, and the energy expectation value of the constrained evolution is already close to the expected result. 
In \isotope[6]{He} the difference between the constrained and unconstrained energy is of 
the order of $1\,\rm MeV$, roughly $3\%$ of the final result. 
In \isotope[16]{O} instead, the constrained energy is higher, 
and the unconstrained evolution lowers 
its value by about $25\,\rm MeV$, $\approx22\%$ of the total energy.
This could be improved by employing more sophisticated wave functions including higher order
correlations, such as in~\cref{eq:psi2}, 
and/or using more refined techniques to perform the unconstrained evolution. 
Studies along these directions are underway.

\subsection{Test of quadratic two-body correlations}
\label{sec:psi2}
The results presented in the previous section are obtained using a trial wave function 
of the form of~\cref{eq:psi}, i.e., by retaining only two-body linear correlations in $\langle RS|\Psi\rangle$.
We present in~\cref{tab:psi2} a test study on the effect of including quadratic correlations
in the wave function on the energy expectation value. The energy expectation values for the constrained evolution have been
calculated for \isotope[4]{He}, \isotope[16]{O}, and symmetric nuclear matter (SNM) with 28 particles 
in a box with periodic boundary conditions at saturation density $\rho_0=0.16$ fm$^{-3}$.
We use the AV6$'$ potential with no Coulomb interaction for all the systems.
Results are shown for the linear, independent pair, and full quadratic two-body correlations.

\begin{table}[htb]
\centering
\caption[]{Energy per nucleon (in MeV) for \isotope[4]{He}, \isotope[16]{O}, and SNM at $\rho_0$.
The employed potential is AV6$'$. No Coulomb interaction is considered here.
Results are shown for the linear, independent pair, and full quadratic two-body correlations.
Errors are statistical.}
\begin{tabular}{ccccc}
\hline\hline
System & Linear & Ind-Pair & Quadratic \\
\hline
\isotope[4]{He} & $-6.79(1) $ & $-6.81(1) $ & $-6.78(1)  $ \\
\isotope[16]{O} & $-7.23(6) $ & $-7.59(9) $ & $-7.50(9)  $ \\
SNM             & $-13.92(6)$ & $-14.80(7)$ & $-14.70(11)$ \\
\hline\hline
\end{tabular}
\label{tab:psi2}
\end{table}

Though there is little difference in \isotope[4]{He}, the constrained energies 
for both \isotope[16]{O} and SNM are lower when employing quadratic correlations, 
particularly for SNM. In \isotope[16]{O} the energy gain for the constrained
evolution is only $\approx0.3(1)\,\rm MeV/A$, while in SNM this value increases
up to $\approx0.8(1)\,\rm MeV/A$.
Within statistical uncertainties, no difference in the results is found between 
independent pair and full quadratic correlations, though the latter have a higher 
computational cost.
Note that the variational parameters in the trial wave function of~\cref{eq:psi2}
were re-optimized for \isotope[4]{He}. In the case of \isotope[16]{O} and SNM instead, 
due to the cost of optimizing such parameters using the full wave function of~\cref{eq:psi2},
we used the same parameters obtained for the linear wave function of~\cref{eq:psi}.

\subsection{Fit of the three-body interaction}
The three-body interaction, which appears naturally in the chiral
expansion at N$^2$LO, introduces two additional LECs that need to be fit
to experimental data.
The choice considered here is to fit the LECs $c_D$ and $c_E$,
multiplying the intermediate- and short-range parts of the $3N$
interaction respectively (see~\cref{eq:lecs}), to two uncorrelated
observables as in Ref.~\cite{Lynn:2016}: the binding energy of \isotope[4]{He} and $n$-$\alpha$
scattering $P$-wave phase shifts.
This choice probes properties of light nuclei (the \isotope[4]{He}
binding energy) while also providing a handle on spin-orbit splitting via
the splitting in the two $P$-wave $n$-$\alpha$ phase shifts.
Furthermore, the $n$-$\alpha$ system is the lightest nuclear system
presenting three interacting neutrons.
It follows that this choice constrains $c_D$ and $c_E$ well,
and also probes $T=3/2$ physics.

\begin{figure}[b]
\includegraphics[width=\linewidth]{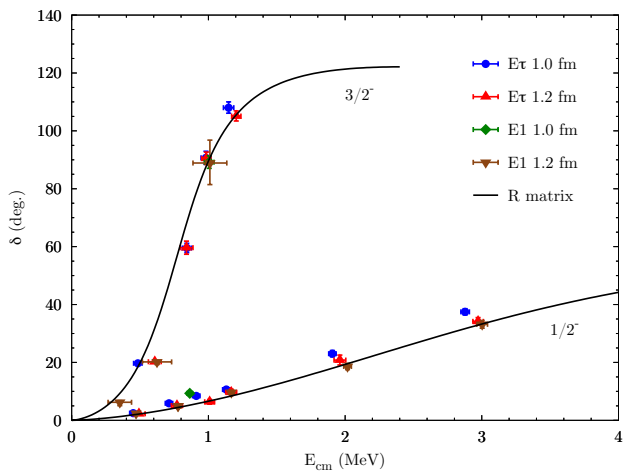}
\caption[]{$P$-wave $n$-$\alpha$ elastic scattering phase shifts compared to an $R$-matrix analysis
of experimental data~\cite{Hale}.}
\label{fig:nalpha}
\end{figure}

The detailed fitting procedure is reported in Ref.~\cite{Lynn:2016},
where different parametrizations of the three-body force for different
cutoffs were explored.
No fit for the $E\mathbbm1$ parametrization and the softer cutoff
$R_0=1.2\,\rm fm$ was reported at that time.
However, in Ref.~\cite{Lonardoni:2017afdmc} a significant overbinding of
\isotope[16]{O} was found for this softer cutoff and the $E\tau$
parametrization of the $3N$ interaction.
Locally regulated chiral interactions spoil the Fierz ambiguity
typically exploited to allow the selection of one of six operators in
the contact interaction $V_E$: see Refs.~\cite{Lynn:2016,Huth:2017} for
details.
This means that observables will depend on the parametrization of the
$3N$ interaction and as suggested in Ref.~\cite{Lynn:2016}, this is
especially true for larger or more dense nuclear systems.
Ref.~\cite{Lynn:2016} also showed that the $E\tau$ parametrization was
the most attractive of the two parametrizations, while the $E\mathbbm1$
parametrization was the least attractive.
Therefore, it became important to consider now the $E\mathbbm1$
parametrization with the softer cutoff $R_0=1.2\,\rm fm$.
This combination is thus explored in this work, together with the
$E\mathbbm1$ parametrization for the $R_0=1.0\,\rm fm$ cutoff, and the
$E\tau$ parametrization for both cutoffs.
In~\cref{fig:nalpha} we report the $P$-wave $n$-$\alpha$ phase shifts
for the four different combinations of operator structure and cutoff
considered in this work.
The corresponding values of $c_D$ and $c_E$ are shown
in~\cref{tab:3bfit}.

\begin{table}[tb]
\centering
\caption[]{LECs $c_D$ and $c_E$ for different cutoffs and parametrizations of the $3N$ force.}
\begin{tabular}{cccc}
\hline\hline
$3N$ & $R_0\,(\rm fm)$ & $c_D$ & $c_E$ \\
\hline
$E\tau$      & $1.0$ & $0.0$   & $-0.63$ \\
             & $1.2$ & $3.5$   & $0.09$  \\
$E\mathbbm1$ & $1.0$ & $0.5$   & $0.62$  \\
             & $1.2$ & $-0.75$ & $0.025$ \\
\hline\hline
\end{tabular}
\label{tab:3bfit}
\end{table}

\subsection{Test of the three-body calculation}
The energies reported in~\cref{fig:tr_he6,fig:tr_o16} correspond to the
expectation values of the effective Hamiltonian $H'$, \cref{eq:h'}. 
These need to be adjusted with the perturbative correction of~\cref{eq:<h3b>}---also extracted 
from the unconstrained evolution---in order to obtain the final results reported in~\cref{tab:10,tab:12}. 
Once the optimal set of parameters $\alpha_i$ is found, these corrections are small,
almost consistent with zero within Monte Carlo statistical uncertainties,
as shown in~\cref{tab:pert}.

\setlength{\tabcolsep}{4pt}
\begin{table}[htb]
\centering
\caption[]{Energy expectation values of~\cref{eq:<h3b>} for $A\ge6$. 
Errors are statistical. Results are in MeV.}
\begin{tabular}{lccccc}
\hline\hline
$\isotope[A]{Z}\,(J^\pi,T)$ & $3N$ & $R_0\,(\rm fm)$ & $\langle H'\rangle$ & $\langle V_{\rm pert}\rangle$ & $\langle H\rangle$ \\
\hline
\isotope[6]{He}\,$(0^+,1)$	& $E\tau$      & $1.0$ & $-28.3(4)$ & $0.1(2)$  & $-28.4(4)$ \\
							&              & $1.2$ & $-29.1(1)$ & $0.2(1)$  & $-29.3(1)$ \\
							& $E\mathbbm1$ & $1.0$ & $-28.5(5)$ & $-0.3(2)$ & $-28.2(5)$ \\
							&              & $1.2$ & $-27.3(3)$ & $0.1(2))$ & $-27.4(4)$ \\
\hline
\isotope[6]{Li}\,$(1^+,0)$	& $E\tau$      & $1.0$ & $-31.2(4)$ & $0.3(3)$  & $-31.5(5)$ \\
							&              & $1.2$ & $-31.9(3)$ & $0.4(1)$  & $-32.3(3)$ \\
							& $E\mathbbm1$ & $1.0$ & $-30.9(4)$ & $-0.2(2)$ & $-30.7(4)$ \\
							&              & $1.2$ & $-30.0(3)$ & $-0.1(2)$ & $-29.9(4)$ \\
\hline
\isotope[12]{C}\,$(0^+,0)$	& $E\tau$      & $1.0$ & $-75(2)$   & $3(1)$    & $-78(3)$   \\
\hline                                     
\isotope[16]{O}\,$(0^+,0)$	& $E\tau$      & $1.0$ & $-115(5)$  & $2(1)$    & $-117(5)$  \\
							&              & $1.2$ & $-265(25)$ & $-2(6)$   & $-263(26)$ \\
							& $E\mathbbm1$ & $1.0$ & $-114(6)$  & $1(2)$    & $-115(6)$  \\
							&              & $1.2$ & $-113(5)$  & $-2(2)$   & $-111(5)$  \\
\hline\hline
\end{tabular}
\label{tab:pert}
\end{table}
\setlength{\tabcolsep}{10pt}

The final result $\langle H\rangle$ is, however, nearly independent of variations
of the $\alpha_i$ parameters, even for larger systems. This is shown in~\cref{tab:alpha} 
where the $\alpha_i$ are arbitrarily changed in \isotope[16]{O} within $5-10\%$ with respect 
to the optimal values, given in the first row for each cutoff. 
This results in $\lesssim4\%$ variations of the total energy, 
compatible with the overall Monte Carlo statistical uncertainties. 
Note that, in order to save computing time, this test has been done using the 
constrained evolution. However, the optimal constrained expectation values 
$\langle V_{\rm pert}\rangle$ are consistent with the unconstrained ones of~\cref{tab:pert}.

\setlength{\tabcolsep}{1.5pt}
\begin{table}[htb]
\centering
\caption[]{Contributions to the energy expectation value of~\cref{eq:<h3b>} in \isotope[16]{O}.
The parametrization $E\tau$ of the $3N$ force is used for different cutoffs.
$\langle V_{\rm pert}\rangle$ is extracted from a mixed estimate, as in~\cref{eq:mix}.
For each cutoff, the first line represents the optimal choice for $\alpha_i$.
Energies (in MeV) are the result of the constrained evolution.
Errors are statistical.}
\begin{tabular}{ccccc}
\hline\hline
$R_0\,(\rm fm)$ & $(\alpha_1,\alpha_2,\alpha_3)$ & $\langle H'\rangle$ & $\langle V_{\rm pert}\rangle$ & $\langle H\rangle$ \\
\hline
$1.0$ & $(2.05,-3.80,-0.95)$ & $-90.0(3)$  & $1.8(5)$    & $-91.8(6)$   \\
      & $(2.50,-3.30,-1.20)$ & $-125.1(6)$ & $-33.9(8)$  & $-92.2(1.0)$ \\
      & $(1.95,-4.00,-0.90)$ & $-83.3(2)$  & $5.9(9)$    & $-89.2(1.0)$ \\
      & $(1.80,-4.20,-0.85)$ & $-75.6(3)$  & $13.9(1.4)$ & $-89.4(1.5)$ \\
\hline                        
$1.2$ & $(1.80,0.45,8.00)$   & $-171(2)$   & $-2(1)$     & $-169(2)$    \\
      & $(1.90,0.50,8.50)$   & $-197(3)$   & $-25(2)$    & $-172(3)$    \\
      & $(1.70,0.40,7.50)$   & $-147(1)$   & $15(1)$     & $-162(1)$    \\
\hline\hline
\end{tabular}
\label{tab:alpha}
\end{table}
\setlength{\tabcolsep}{10pt}

Unless specified otherwise, in the following, all ground-state energies  
correspond to the final expectation value $\langle H\rangle$, extracted from 
the unconstrained Monte Carlo results for $\langle H'\rangle$ with an exponential fit, 
and adjusted with the perturbative correction of~\cref{eq:<h3b>} when $3N$ forces are employed.

\subsection{Ground-state energies and charge radii}
We consider local chiral Hamiltonians at leading-order (LO), 
next-to-leading-order (NLO), and N$^2$LO, the latter including both two- and three-body forces.
At each order we can assign theoretical uncertainties to observables coming from the 
truncation of the chiral expansion, see, e.g., Ref.~\cite{Epelbaum:2015epja}. 
For an observable $X$ at N$^2$LO, 
the theoretical uncertainty is obtained as
\begin{align}
\Delta X^{\text{N}^2\text{LO}}=&\max(Q^4\times|X^{\text{LO}}|,\nonumber \\ 
&\phantom{\max(\,}Q^2\times|X^{\text{NLO}}-X^{\text{LO}}|, \nonumber \\
&\phantom{\max(\,}Q^{\phantom{2}}\times|X^{\text{N}^2\text{LO}}-X^{\text{NLO}}|),
\label{eq:err}
\end{align}
where we take $Q=m_\pi/\Lambda_b$ with $m_\pi\approx140\,\rm MeV$ and $\Lambda_b=600$~MeV, 
as in Ref.~\cite{Lonardoni:2017afdmc}.

The expectation value of the charge radius is derived from the point-proton radius using the relation:
\begin{align}
	\left\langle r_{\rm ch}^2\right\rangle=
	\left\langle r_{\rm pt}^2\right\rangle+
	\left\langle R_p^2\right\rangle+
	\frac{A-Z}{Z}\left\langle R_n^2\right\rangle+
	\frac{3\hbar^2}{4M_p^2c^2},
	\label{eq:rch}
\end{align}
where $r_{\rm pt}$ is the calculated point-proton radius,
$\left\langle R_p^2\right\rangle=0.770(9)\,\rm{fm}^2$~\cite{Beringer:2012} the proton radius, 
$\left\langle R_n^2\right\rangle=-0.116(2)\,\rm{fm}^2$~\cite{Beringer:2012} the neutron radius,
and $(3\hbar^2)/(4M_p^2c^2)\approx0.033\,\rm{fm}^2$ the Darwin-Foldy correction~\cite{Friar:1997}.
For \isotope[6]{He} a spin-orbit correction 
$\left\langle r_{\rm so}^2\right\rangle=-0.08\,\rm{fm}^2$~\cite{Ong:2010} is also included.
The point-nucleon radius $r_{\rm pt}$ is calculated as
\begin{align}
	\left\langle r_N^2\right\rangle=\frac{1}{{\cal N}}\big\langle\Psi\big|\sum_i\mathcal P_{N_i} |\vb{r}_i-\vb{R}_{\rm cm}|^2\big|\Psi\big\rangle,
\end{align}
where $\vb{R}_{\rm cm}$ is the coordinate of the center of mass of the system,
${\cal N}$ is the number of protons or neutrons, and 
\begin{align}
	\mathcal P_{N_i}=\frac{1\pm\tau_{z_i}}{2},
	\label{eq:proj}
\end{align}
is the projector operator onto protons or neutrons. 
The charge radius is a mixed expectation value, and it requires the calculation of both
VMC and DMC point-proton radii, according to~\cref{eq:mix}. 
Regardless of the employed optimization of the variational wave function (free or constrained), 
the extrapolation of the mixed estimate $\left\langle r_{\rm ch}^2\right\rangle$ is small, 
and the final results for different optimizations typically agree within statistical uncertainties.

The ground-state energies and charge radii for light systems $(A=3,4)$ employing 
local chiral potential at N$^2$LO are shown in~\cref{tab:afdmc-gfmc}. 
Results with ($E\tau$ parametrization) and without the $3N$ force are shown 
for different choices of the cutoff $R_0$. For all the $s_{1/2}$, systems we used the same 
parameters $\alpha_i$ for the propagation of the $3N$ force, 
determined in order to minimize the perturbative correction of~\cref{eq:<h3b>}.
The agreement with the GFMC results of Ref.~\cite{Lynn:2016,Lynn:2017},
where the $3N$ interactions are fully included in the propagation, 
is within a few percent both at the two- and three-body level, providing a good
benchmark for the AFDMC propagation technique described in~\cref{sec:p3}.

\begin{table*}[htb]
\centering
\caption[]{Ground-state energies and charge radii for $A=3,4$ employing local chiral potentials at N$^2$LO. 
The $E\tau$ parametrization of the $3N$ force is used. Errors are statistical. 
GFMC results are from Refs.~\cite{Lynn:2014,Lynn:2016}.}
\begin{tabular}{cclcccc}
\hline\hline
Nucleus & Cutoff & Potential & \multicolumn{2}{c}{AFDMC} & \multicolumn{2}{c}{GFMC} \\
$\isotope[A]{Z}\,(J^\pi,T)$ & $R_0\,(\rm fm)$ & & $E\,(\rm MeV)$ & $r_{\rm ch}\,(\rm fm)$ & $E\,(\rm MeV)$ & $r_{\rm ch}\,(\rm fm)$ \\     
\hline
\isotope[3]{H}\,$(\frac{1}{2}^+,\frac{1}{2})$  & $1.0$ & $N\!N$       & $-7.54(4)$   & $1.75(2)$ & $-7.55(1)$  & $1.78(2)$ \\
   	  	                                       &       & $3N$ $E\tau$ & $-8.33(7)$   & $1.72(2)$ & $-8.34(1)$  & $1.72(3)$ \\
                                               & $1.2$ & $N\!N$       & $-7.76(3)$   & $1.74(2)$ & $-7.74(1)$  & $1.75(2)$ \\
   	  	                                       &       & $3N$ $E\tau$ & $-8.27(5)$   & $1.73(2)$ & $-8.35(4)$  & $1.72(4)$ \\
\hline                                                               
\isotope[3]{He}\,$(\frac{1}{2}^+,\frac{1}{2})$ & $1.0$ & $N\!N$       & $-6.89(5)$   & $2.02(2)$ & $-6.78(1)$  & $2.06(2)$ \\
                                               &       & $3N$ $E\tau$ & $-7.55(8)$   & $1.96(2)$ & $-7.65(2)$  & $1.97(2)$ \\
                                               & $1.2$ & $N\!N$       & $-7.12(3)$   & $1.98(2)$ & $-7.01(1)$  & $2.01(1)$ \\
                                               &       & $3N$ $E\tau$ & $-7.64(4)$   & $1.95(5)$ & $-7.63(4)$  & $1.97(1)$ \\
\hline                                                        
\isotope[4]{He}\,$(0^+,0)$                     & $1.0$ & $N\!N$       & $-23.96(8)$  & $1.72(2)$ & $-23.72(1)$ & $1.73(1)$ \\
                                               &       & $3N$ $E\tau$ & $-27.64(13)$ & $1.68(2)$ & $-28.30(1)$ & $1.65(2)$ \\
                                               & $1.2$ & $N\!N$       & $-25.17(5)$  & $1.69(1)$ & $-24.86(1)$ & $1.69(1)$ \\
                                               &       & $3N$ $E\tau$ & $-28.37(8)$  & $1.65(1)$ & $-28.30(1)$ & $1.64(1)$ \\
\hline\hline
\end{tabular}
\label{tab:afdmc-gfmc}
\end{table*}

\begin{figure*}[htb]
\includegraphics[width=0.48\linewidth]{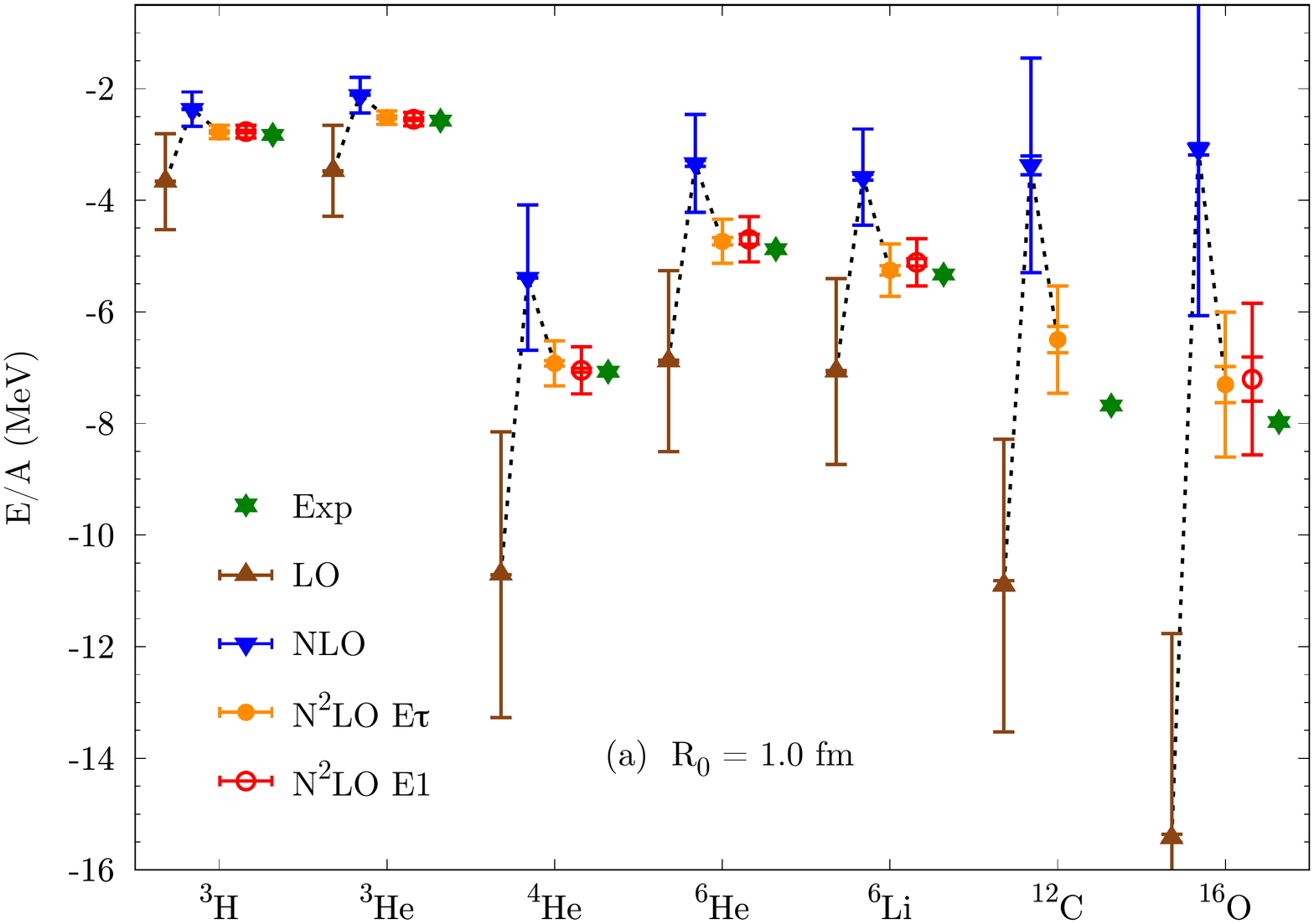}\qquad\includegraphics[width=0.48\linewidth]{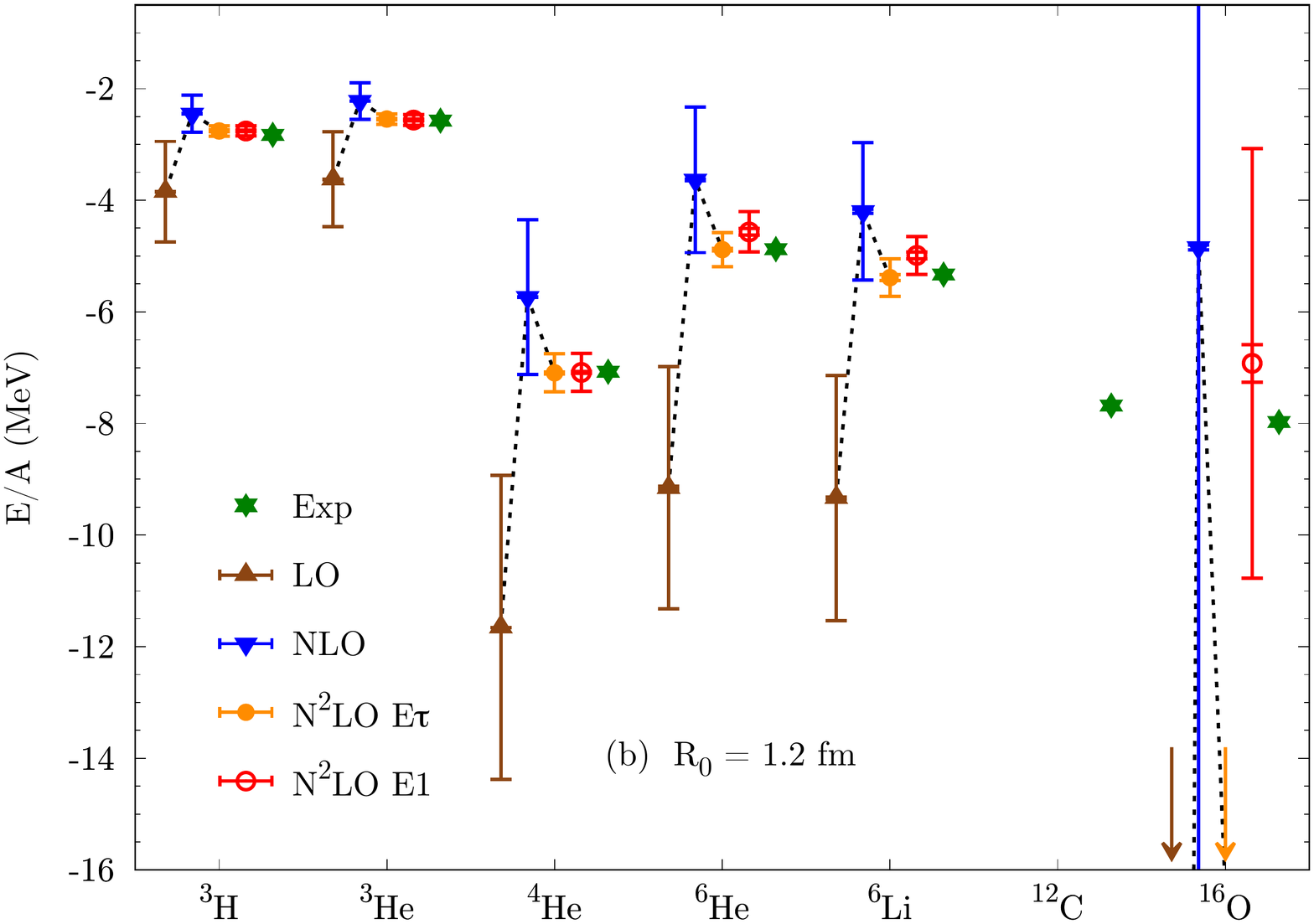}
\caption[]{Ground-state energies per nucleon for $3\le A\le16$ with local chiral potentials:
(a) $R_0=1.0\,\rm fm$ cutoff (left panel), (b) $R_0=1.2\,\rm fm$ cutoff (right panel). 
Results at different orders of the chiral expansion and for different $3N$ parametrizations are shown.
Smaller error bars (indistinguishable from the symbols up to $A=6$) indicate the statistical Monte Carlo uncertainty,
while larger error bars are the uncertainties from the truncation of the chiral expansion. 
LO and N$^2$LO $E\tau$ results for \isotope[16]{O} with $R_0=1.2\,\rm fm$ are outside the displayed energy region.
Updated from Ref.~\cite{Lonardoni:2017afdmc}.}
\label{fig:ene}
\end{figure*}

In~\cref{fig:ene} we present the ground-state energies per nucleon
of nuclei with $3\le A\le16$ for cutoffs $R_0=1.0\,\rm fm$ and $R_0=1.2\,\rm fm$, 
respectively. Results at LO, NLO, and N$^2$LO for both $E\tau$ and $E\mathbbm1$
parametrizations of the $3N$ force are shown. Error bars are estimated by 
including both the Monte Carlo uncertainties and the errors given by 
the truncation of the chiral expansion, the latter being the dominant ones.
For the harder interaction $(R_0=1.0\,\rm fm)$, the predicted binding energies at N$^2$LO 
are in good agreement with experimental data all the way up to $A=16$.
No differences, within theoretical uncertainties, are found for the two different
parametrizations of the $3N$ force.

\isotope[12]{C} in the $E\tau$ parametrization is slightly underbound.
This is most likely a consequence of the employed wave function that results 
in a too high energy for the constrained evolution. 
This could be due to the complicated clustering structure of $^{12}$C not included
in $\Psi_T$, which would require a much longer unconstrained propagation
to filter out the corresponding low excitations from $\Psi_T$.
For $A=6$ the wave
function is constructed using up to $sd$-shell single-particle orbitals. For \isotope[12]{C} 
instead, coupling $p$-shell orbitals only
already results in a sum of 119 Slater determinants. Including orbitals in the $sd$-shell could 
in principle result in a better wave function for this open-shell system, but it will 
sizably increase the number of determinants to consider, making the calculation prohibitively 
time consuming. Another possible improvement would be to include quadratic terms in
the pair correlations, as shown in~\cref{eq:psi2}. However, first attempts in
\isotope[16]{O} lead to just a $\approx6(2)\,\rm MeV$ reduction of the total 
energy in a simplified scenario (see~\cref{tab:psi2}), with a noticeably increased 
computational cost.

For the softer interaction $(R_0=1.2\,\rm fm)$, NLO and in particular LO results are 
typically more bound compared to the $R_0=1.0\,\rm fm$ case. Both parametrizations
of the $3N$ force bring the N$^2$LO energies compatible with the experimental 
values up to $A=6$, and consistent with those obtained with the hard potential. 

\begin{figure*}[htb]
\includegraphics[width=0.48\linewidth]{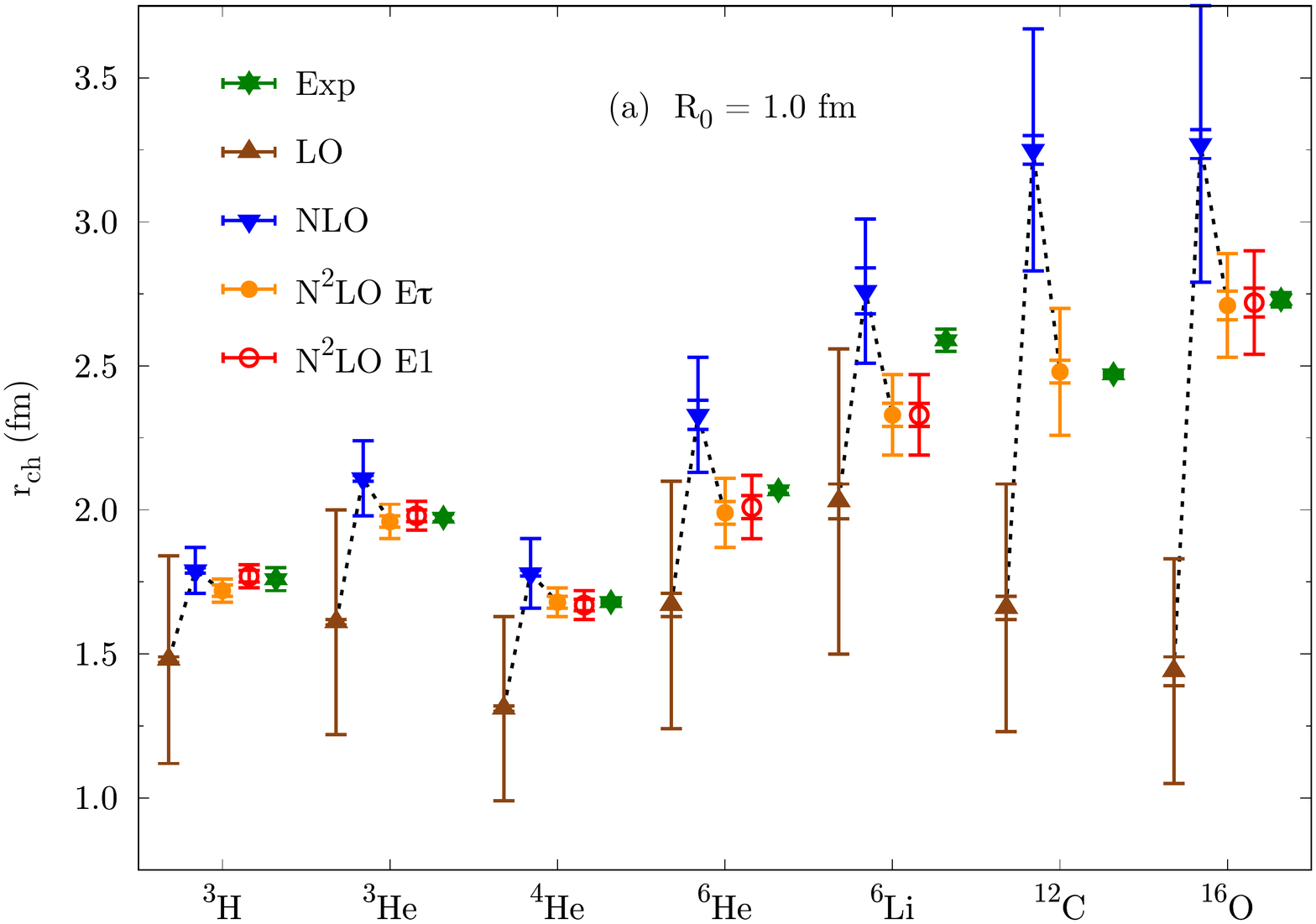}\qquad\includegraphics[width=0.48\linewidth]{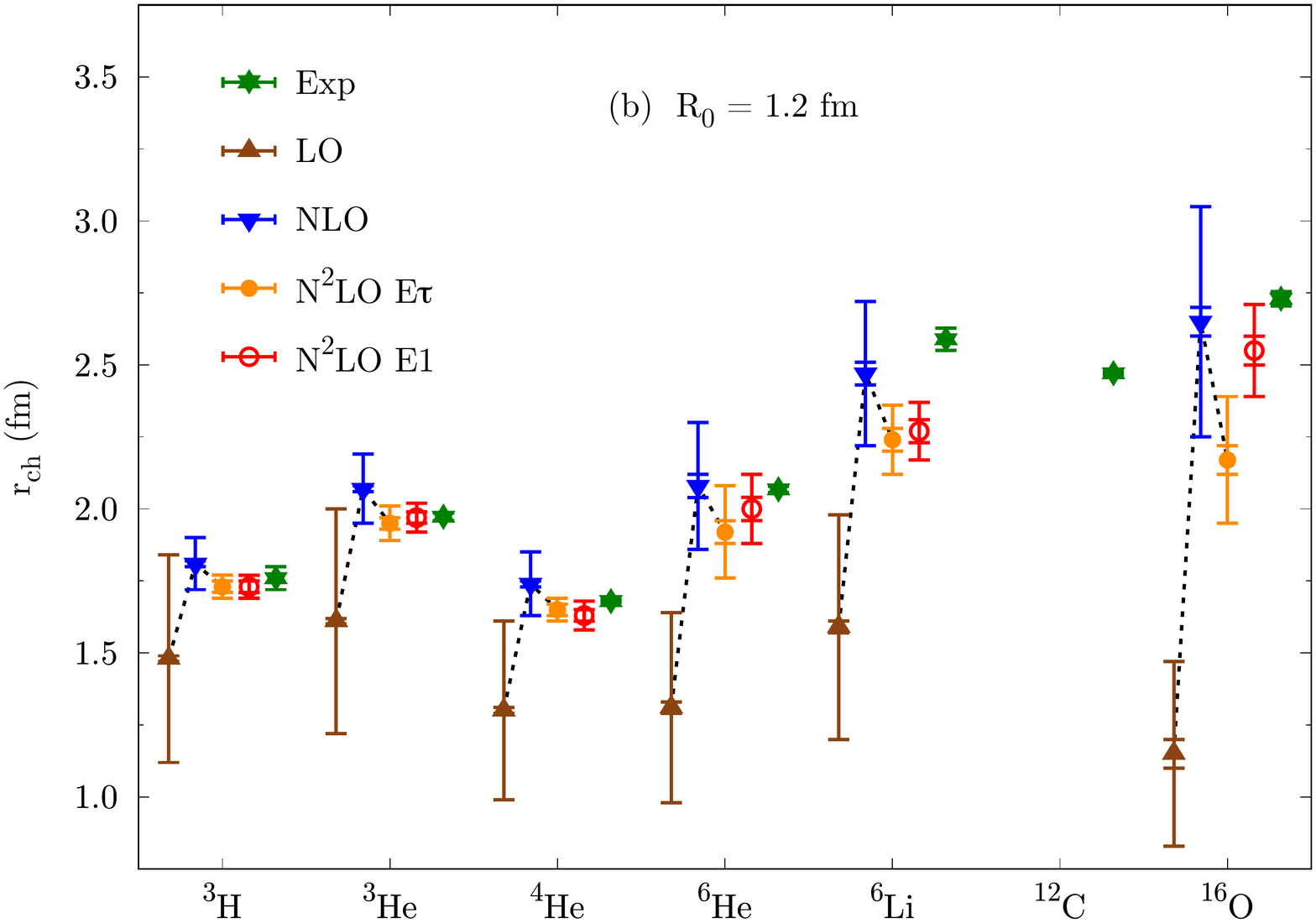}
\caption[]{Charge radii for $3\le A\le16$ with local chiral potentials:
(a) $R_0=1.0\,\rm fm$ cutoff (left panel), (b) $R_0=1.2\,\rm fm$ cutoff (right panel). 
The legend and error bars are as in~\cref{fig:ene}. Updated from Ref.~\cite{Lonardoni:2017afdmc}.}
\label{fig:rch}
\end{figure*}

For the heaviest system considered here, \isotope[16]{O}, the picture is quite different. 
At LO, the system is dramatically overbound $(\approx -1\,\rm GeV)$, which would imply 
very large theoretical uncertainties at NLO and N$^2$LO coming from the prescription of~\cref{eq:err}.
Within these uncertainties, NLO and N$^2$LO two-body energies 
are compatible with the corresponding results for the hard interaction (see~\cref{tab:10,tab:12}).
However, the contribution of the $3N$ force at N$^2$LO largely depends upon
the employed operator structure. The $E\tau$ parametrization for the soft potential is very attractive,
adding almost $10\,\rm MeV$ per nucleon to the total energy, and thus
predicting a significant overbinding with a ground-state energy of $\approx -260\,\rm MeV$.
The $E\mathbbm1$ parametrization is instead less attractive, resulting in $\approx 0.30\,\rm MeV$ per 
nucleon more binding with respect to the two-body case, compatible
with the energy expectation values for the hard potential.

\Cref{fig:rch} shows the charge radii at different orders of the chiral 
expansion and for different cutoffs and parametrizations of the $3N$ force. 
The agreement with experimental data for the hard interaction at N$^2$LO is remarkably
good all the way up to oxygen. One exception is \isotope[6]{Li}, for which the charge radius
is somewhat underpredicted. However, a similar conclusion is found in GFMC calculations
employing the AV18+IL7 potential, where charge radii of lithium isotopes are 
underestimated~\cite{Carlson:2015}. 

For the soft interaction, the description of charge radii resembles order by order 
that for the hard potential up to $A=6$, with the N$^2$LO results in agreement 
with experimental data, except for \isotope[6]{Li} (also shown in~\cref{tab:afdmc-gfmc}). 
The picture changes again for $A=16$. The charge radius of \isotope[16]{O} turns out to 
be close to $2.2\,\rm fm$ with the $E\tau$ parametrization of the $3N$ force, 
smaller than that of \isotope[6]{Li} for the same potential, but consistent with the 
significant overbinding predicted for $A=16$. The oxygen charge radius for the $E\mathbbm1$ 
parametrization is instead closer to the experimental value.

The details of LO, NLO, and N$^2$LO calculations for $A\ge6$ are reported in~\cref{tab:10,tab:12}
for $R_0=1.0\,\rm fm$ and $R_0=1.2\,\rm fm$, respectively. Results for the constrained and 
unconstrained evolution energies are both shown, together with the charge radii.
Both Monte Carlo uncertainties and theoretical errors coming from the truncation of the chiral 
expansion are reported (where available). At N$^2$LO the two-body energy is shown together with
that of the two different parametrizations of the $3N$ force ($E\tau$ and $E\mathbbm{1}$).

The full calculation of \isotope[12]{C} at N$^2$LO required on the
order of $10^6$ CPU hours (on Intel Broadwell cores @ 2.1GHz) for a single cutoff $(1.0\,\rm fm)$ 
and $3N$ parametrization $(E\tau)$. Due to the high computational cost, no attempts were
made for the $E\mathbbm{1}$ parametrization of the $3N$ force or for the $1.2\,\rm fm$ cutoff.

\begin{table*}[htb]
\centering
\caption[]{Ground-state energies and charge radii for $A\ge6$ with local chiral potentials. 
Results at different orders of the chiral expansion and for different $3N$ parametrizations are shown. 
Energy results are shown for both the constrained ($E_{\rm C}$) and unconstrained ($E$) evolutions.
The first error is statistical, the second is based on the EFT expansion uncertainty.
The employed cutoff is $R_0=1.0\,\rm fm$.}
\begin{tabular}{llccc}
\hline\hline
$\isotope[A]{Z}\,(J^\pi,T)$ & Potential & $E_{\rm C}\,(\rm MeV)$ & $E\,(\rm MeV)$ & $r_{\rm ch}\,(\rm fm)$ \\
\hline
\isotope[6]{He}\,$(0^+,1)$ & LO                          & $-42.1(1)$  & $-41.3(1)(9.6)$ & $1.67(4)(39)$ \\
                           & NLO                         & $-18.19(7)$ & $-20.0(3)(5.0)$ & $2.33(5)(15)$ \\
                           & N$^2$LO $N\!N$              & $-22.24(4)$ & $-23.1(2)(1.2)$ & $2.11(4)(5)$  \\
   	  	                   & N$^2$LO $3N$ $E\tau$        & $-26.58(6)$ & $-28.4(4)(2.0)$ & $1.99(4)(8)$  \\
   	  	                   & N$^2$LO $3N$ $E\mathbbm{1}$ & $-26.33(8)$ & $-28.2(5)(1.9)$ & $2.01(4)(7)$  \\
                           & exp                         &             & $-29.3$         & $2.068(11)$~\cite{Mueller:2007} \\
\hline                                              
\isotope[6]{Li}\,$(1^+,0)$ & LO                          & $-42.8(1)$  & $-42.4(1)(9.9)$ & $2.03(6)(47)$ \\
                           & NLO                         & $-19.2(2)$  & $-21.5(3)(4.9)$ & $2.76(8)(17)$ \\
                           & N$^2$LO $N\!N$              & $-24.3(1)$  & $-25.5(4)(1.1)$ & $2.46(4)(7)$  \\
   	  	                   & N$^2$LO $3N$ $E\tau$        & $-28.9(1)$  & $-31.5(5)(2.3)$ & $2.33(4)(10)$ \\
   	  	                   & N$^2$LO $3N$ $E\mathbbm{1}$ & $-28.9(1)$  & $-30.7(4)(2.1)$ & $2.33(4)(10)$ \\
                           & exp                         &             & $-32.0$         & $2.589(39)$~\cite{Nortershauser:2011} \\
\hline                                                   
\isotope[12]{C}\,$(0^+,0)$ & LO                          & $-131.5(2)$ & $-131(1)(31)$   & $1.66(4)(39)$ \\
                           & NLO                         & $-31.1(2)$  & $-41(2)(21)$    & $3.25(5)(37)$ \\
                           & N$^2$LO $N\!N$              & $-63.5(2.4)$& $-66(3)(6)$     & $2.66(4)(14)$ \\
   	  	                   & N$^2$LO $3N$ $E\tau$        & $-70.2(5)$  & $-78(3)(9)$     & $2.48(4)(18)$ \\
   	  	                   & N$^2$LO $3N$ $E\mathbbm{1}$ & $-$         & $-$             & $-$           \\
                           & exp                         &             & $-92.2$         & $2.471(6)$~\cite{Sick:1982} \\
\hline                                                   
\isotope[16]{O}\,$(0^+,0)$ & LO                          & $-251.7(2)$ & $-247(1)(58)$   & $1.44(3)(34)$ \\
                           & NLO                         & $-37.3(2)$  & $-49(2)(46)$    & $3.27(5)(43)$ \\
                           & N$^2$LO $N\!N$              & $-72.8(2)$  & $-87(3)(11)$    & $2.76(5)(12)$ \\
   	  	                   & N$^2$LO $3N$ $E\tau$        & $-91.8(6)$  & $-117(5)(16)$   & $2.71(5)(13)$ \\
   	  	                   & N$^2$LO $3N$ $E\mathbbm{1}$ & $-85.8(5)$  & $-115(6)(15)$   & $2.72(5)(11)$ \\
                           & exp                         &             & $-127.6$        & $2.730(25)$~\cite{Sick:1970} \\
\hline\hline
\end{tabular}
\label{tab:10}
\end{table*}

\begin{table*}[htb]
\centering
\caption[]{Same as~\cref{tab:10} but for the $R_0=1.2\,\rm fm$ cutoff.}
\begin{tabular}{llccc}
\hline\hline
$\isotope[A]{Z}\,(J^\pi,T)$ & Potential & $E_{\rm C}\,(\rm MeV)$ & $E\,(\rm MeV)$ & $r_{\rm ch}\,(\rm fm)$ \\
\hline                                        
\isotope[6]{He}\,$(0^+,1)$ & LO                     & $-55.65(6)$  & $-54.9(2)(12.8)$  & $1.31(2)(31)$ \\
                           & NLO                    & $-21.41(6)$  & $-21.8(1)(7.7)$   & $2.08(4)(18)$ \\
                           & N$^2$LO $N\!N$         & $-24.25(5)$  & $-24.3(1)(1.8)$   & $2.02(4)(4)$  \\
   	  	                   & N$^2$LO $E\tau$        & $-28.37(5)$  & $-29.3(1)(1.8)$   & $1.92(4)(4)$  \\
                           & N$^2$LO $E\mathbbm{1}$ & $-26.98(8)$  & $-27.4(4)(1.8)$   & $2.00(4)(4)$ \\
                           & exp                    &              & $-29.3$           & $2.068(11)$~\cite{Mueller:2007} \\
\hline                                                             
\isotope[6]{Li}\,$(1^+,0)$ & LO                     & $-56.84(3)$  & $-56.0(1)(13.1)$  & $1.59(2)(37)$ \\
                           & NLO                    & $-23.64(8)$  & $-25.2(2)(7.2)$   & $2.47(4)(21)$ \\
                           & N$^2$LO $N\!N$         & $-26.76(3)$  & $-27.0(2)(1.7)$   & $2.41(4)(5)$  \\
   	  	                   & N$^2$LO $E\tau$        & $-30.8(1)$   & $-32.3(3)(1.7)$   & $2.24(4)(6)$  \\
                           & N$^2$LO $E\mathbbm{1}$ & $-29.2(1)$   & $-29.9(4)(1.7)$   & $2.29(4)(5)$ \\
                           & exp                    &              & $-32.0$           & $2.589(39)$~\cite{Nortershauser:2011} \\
\hline                                       
\isotope[16]{O}\,$(0^+,0)$ & LO                     & $-1158.8(5)$ & $-1110(31)(259)$  & $1.15(5)(27)$ \\
                           & NLO                    & $-72.3(1)$   & $-77.5(7)(240.8)$ & $2.65(5)(35)$ \\
                           & N$^2$LO $N\!N$         & $-98.6(1)$   & $-106(4)(56)$     & $2.47(5)(8)$  \\
   	  	                   & N$^2$LO $E\tau$        & $-169(2)$    & $-263(26)(56)$    & $2.17(5)(11)$ \\
                           & N$^2$LO $E\mathbbm{1}$ & $-99.5(4)$   & $-111(5)(56)$     & $2.55(5)(8)$ \\
                           & exp                    &              & $-127.6$          & $2.730(25)$~\cite{Sick:1970} \\
\hline\hline
\end{tabular}
\label{tab:12}
\end{table*}

As shown in~\cref{tab:10,tab:12}, the overbinding in \isotope[16]{O} happens only when the $3N$ force 
is included with the $E\tau$ parametrization for $R_0=1.2\,\rm fm$. The alternative combinations of three-body 
operators and cutoffs considered in this work predict instead binding energies compatible with the 
experimental value. A close look at the energy contributions coming from the $3N$ force in 
\isotope[6]{Li} and \isotope[16]{O} (\cref{tab:v3}) clearly shows the issue. 
A large negative $V_D$ contribution in \isotope[16]{O} for the soft $E\tau$ potential drives the system to 
a strongly bound state. In fact, even though the total energy at the two-body level is similar 
to that of the other soft potentials for $A=16$, the individual expectation values for the kinetic
energy and the two-body potential are severely larger, consistent with a very compact system. 
The $3N$ force adds then $\approx 13\,\rm MeV$ per nucleon, roughly half coming from the also
increased TPE contribution, and half from $V_D$. In the case of the $R_0=1.0\,\rm fm$ cutoff instead, 
the $3N$ force in both parametrizations adds only $<3\,\rm MeV$ per nucleon to the total
two-body energy, with similar TPE contributions and a balance between $\langle V_D\rangle$ and
$\langle V_E\rangle$. This is still true in \isotope[6]{Li} also for $R_0=1.2\,\rm fm$, but 
the balance is broken for the soft $E\tau$ potential in \isotope[16]{O}. The main reason for such
behavior can be attributed to the large value of $c_D$ for this potential 
(see~\cref{tab:3bfit}), particularly effective for $A>6$. 

As has been discussed briefly above and in more detail in Refs.~\cite{Lynn:2016,Huth:2017},
locally regulated chiral interactions spoil the Fierz rearrangement
freedom used to select one of the six possible operators that are
consistent with the symmetries of the theory for the contact interaction
at N$^2$LO, $V_E$. This means that results at finite cutoff depend on
this choice. However, these additional regulator artifacts are absorbed
by higher-order LECs in chiral EFT~\cite{Huth:2017}. Furthermore, the dependence is
typically within the truncation uncertainties (an exception occurs for
denser or heavier systems such as neutron matter beyond saturation
density, or as shown above, \isotope[16]{O}). In these cases, since the
next order in chiral EFT where $3N$ contacts appear is
next-to-next-to-next-to-next-to-leading order, a significant challenge
at this point, one can use instead the parametrization
$V_{E\mathcal{P}}$ of the contact interaction introduced in Ref.~\cite{Lynn:2016},
which projects onto the total spin $S=1/2$ and total isospin $T=1/2$
triples. These are the triples that survive in the infinite
(momentum-space) cutoff limit and thus, this parametrization partially
restores the Fierz rearrangement freedom. However, the
$V_{E\mathcal{P}}$ parametrization involves spin/isospin operators
beyond quadratic order and presents a challenge to the direct inclusion
in the AFDMC propagation. We leave the exploration of this
parametrization to future works.

\setlength{\tabcolsep}{5pt}
\begin{table*}[htb]
\centering
\caption[]{Expectation value of the N$^2$LO energy contributions in \isotope[6]{Li} and \isotope[16]{O}.
All energies (in MeV) are mixed estimates from the constrained evolution: 
$2\,\langle\mathcal O_{\rm DMC}\rangle - \langle\mathcal O_{\rm VMC}\rangle$.
Errors are statistical.}
\begin{tabular}{cclcccccccc}
\hline\hline
System & $R_0\,(\rm fm)$ & Potential & $E_{\rm kin}$ & $v_{ij}$ & $E_{\rm kin}+v_{ij}$ & $V_{ijk}$ & $V^{2\pi,P}$ & $V^{2\pi,S}$ & $V_D$ & $V_{E}$ \\
\hline                                                                                                                                                         
\isotope[6]{Li} & 1.0 & $N\!N$            & $116.8(4)$ & $-151.2(4)$ & $-34.4(8)$   &            &            &            &            &            \\
                & 1.0 & $3N$ $E\tau$      & $135.3(7)$ & $-165.6(5)$ & $-30.2(1.2)$ & $-11.1(3)$ & $-13.3(3)$ & $-0.43(1)$ & $0$        & $2.67(2)$  \\
                & 1.0 & $3N$ $E\mathbbm1$ & $135.5(6)$ & $-165.8(6)$ & $-30.3(1.2)$ & $-11.3(2)$ & $-13.3(2)$ & $-0.42(1)$ & $-0.89(2)$ & $3.38(4)$  \\ [0.2cm]
                & 1.2 & $N\!N$            & $110.3(3)$ & $-145.4(3)$ & $-35.1(6)$   &            &            &            &            &            \\
                & 1.2 & $3N$ $E\tau$      & $129.3(6)$ & $-160.1(5)$ & $-30.8(1.1)$ & $-11.8(3)$ & $-6.1(2)$  & $-0.39(1)$ & $-4.6(1)$  & $-0.63(1)$ \\
                & 1.2 & $3N$ $E\mathbbm1$ & $118.8(4)$ & $-154.0(3)$ & $-35.2(7)$   & $-5.5(1)$  & $-5.6(1)$  & $-0.26(1)$ & $0.08(1)$  & $0.27(1)$  \\
\hline
\isotope[16]{O} & 1.0 & $N\!N$            & $319(1)$   & $-453(1)$   & $-134(2)$    &            &            &            &            &            \\
                & 1.0 & $3N$ $E\tau$      & $370(1)$   & $-500(1)$   & $-130(2)$    & $-44(1)$   & $-55(1)$   & $0.85(1)$  & $0$        & $8.50(4)$  \\
                & 1.0 & $3N$ $E\mathbbm1$ & $367(1)$   & $-497(1)$   & $-131(2)$    & $-41(1)$   & $-54(1)$   & $0.72(1)$  & $-4.03(5)$ & $15.7(1)$  \\ [0.2cm]
                & 1.2 & $N\!N$            & $377(1)$   & $-528(2)$   & $-151(3)$    &            &            &            &            &            \\
                & 1.2 & $3N$ $E\tau$      & $556(4)$   & $-712(3)$   & $-156(7)$    & $-202(3)$  & $-101(2)$  & $-0.72(9)$ & $-94(2)$   & $-5.43(3)$ \\
                & 1.2 & $3N$ $E\mathbbm1$ & $377(1)$   & $-529(1)$   & $-152(2)$    & $-26(1)$   & $-34(1)$   & $0.94(1)$  & $4.53(8)$  & $1.90(1)$  \\
\hline\hline
\end{tabular}
\label{tab:v3}
\end{table*}
\setlength{\tabcolsep}{10pt}

\subsection{Charge form factors and Coulomb sum rules}
One- and two-body point-nucleon densities are calculated as
\begin{align}
	\!\!\rho_{N}(r) &=\frac{1}{4\pi r^2}\big\langle\Psi\big|\sum_i    \mathcal P_{N_i}\delta(r-|\vb{r}_i-\vb{R}_{\rm cm}|)\big|\Psi\big\rangle, \label{eq:rho_N} \\
	\!\!\rho_{NN}(r)&=\frac{1}{4\pi r^2}\big\langle\Psi\big|\sum_{i<j}\mathcal P_{N_i}P_{N_j}\delta(r-|\vb{r}_i-\vb{r}_j|)\big|\Psi\big\rangle, \label{eq:rho_NN}
\end{align}
where $\mathcal P_{N_i}$ is the projector operator of~\cref{eq:proj}.
With the current definitions, $\rho_N$ and $\rho_{NN}$ integrate to the number 
of nucleons and the number of nucleon pairs, respectively.

As opposed to the charge radius, densities are not observables themselves, 
but the one-body densities can be related to physical quantities experimentally accessible via 
electron-nucleon scattering processes, such as the longitudinal elastic (charge) form factor.
In fact, the charge form factor can be expressed as the ground-state expectation value of 
the one-body charge operator~\cite{Mcvoy:1962},
which, ignoring small spin-orbit contributions in the one-body current, 
results in the following expression:
\begin{align}
	F_L(q)=\frac{1}{Z}\frac{G_E^p(Q_{\rm el}^2)\,\tilde{\rho}_p(q)+G_E^n(Q_{\rm el}^2)\,\tilde{\rho}_n(q)}{\sqrt{1+Q_{\rm el}^2/(4 m_N^2)}},
	\label{eq:ff}
\end{align}
where $\tilde{\rho}_{N}(q)$ is the Fourier transform of the one-body point-nucleon density defined in~\cref{eq:rho_N}, 
and $Q^2_{\rm el}=\vb{q}^2-\omega_{\rm el}^2$ is the four-momentum squared, 
with $\omega_{\rm el}=\sqrt{q^2+m_A^2}-m_A$ the energy transfer corresponding to the elastic peak,
$m_A$ being the mass of the target nucleus.
$G_E^N(Q^2)$ are the nucleon electric form factors, for which we adopt Kelly's parametrization~\cite{Kelly:2004}.

\begin{figure}[htb]
\includegraphics[width=\linewidth]{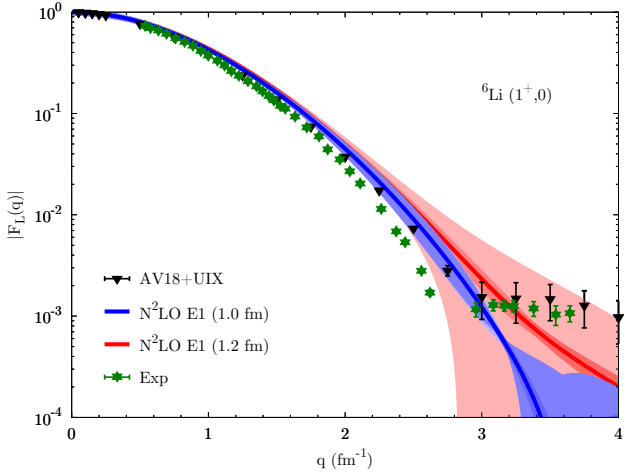}
\caption[]{Charge form factor in \isotope[6]{Li}. The solid blue (red) line is the AFDMC result for the
N$^2$LO $E\mathbbm1$ potential with cutoff $R_0=1.0\,(1.2)\,\rm fm$.
Lighter shaded areas indicate the uncertainties from the truncation of the chiral expansion.
Darker shaded areas are the theoretical error bands only taking into account NLO and N$^2$LO results. 
Black triangles are the VMC one-body results for AV18+UIX~\cite{Wiringa:1998}.
The experimental data are taken from Ref.~\cite{Li:1971}.}
\label{fig:ff_li6}
\end{figure}

\begin{figure}[b]
\includegraphics[width=\linewidth]{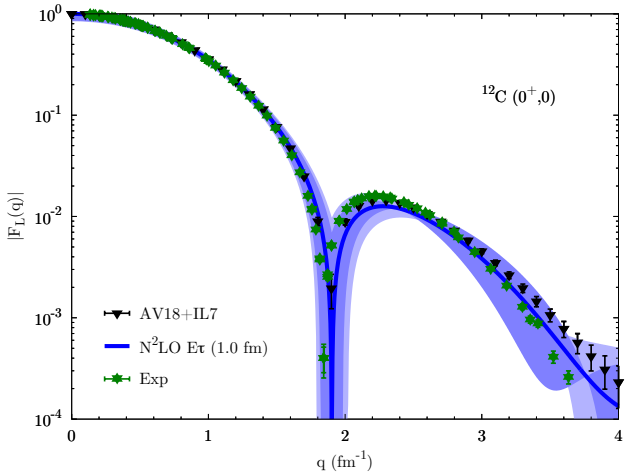}
\caption[]{Charge form factor in \isotope[12]{C}. 
In blue are the AFDMC results for the $E\tau$ parametrization of the $3N$
force and cutoff $R_0=1.0\,\rm fm$.
Black triangles are the GFMC one-body results for AV18+IL7~\cite{Lovato:2013}.
The experimental data are taken from Ref.~\cite{Devries:1987}.
Updated from Ref.~\cite{Lonardoni:2017afdmc}.}
\label{fig:ff_c12}
\end{figure}

\begin{figure}[htb]
\includegraphics[width=\linewidth]{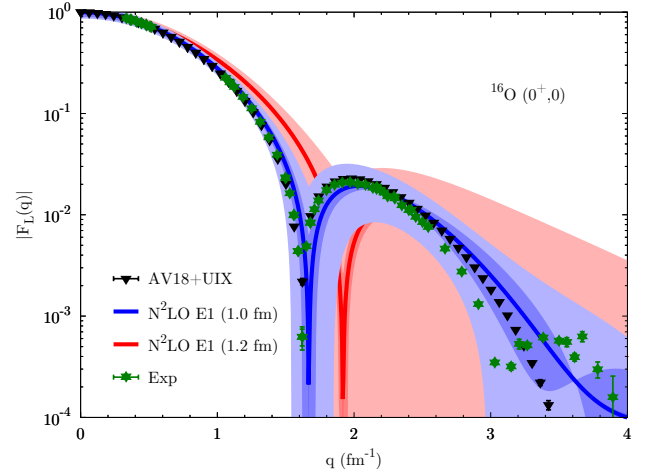}
\caption[]{Charge form factor in \isotope[16]{O}.
In blue (red) are the AFDMC results as in~\cref{fig:ff_li6}.
Black triangles are the cluster-VMC one-body results for AV18+UIX~\cite{Lonardoni:2017cvmc}.
Experimental data are from I. Sick, based on Refs.~\cite{Sick:1970,Schuetz:1975,Sick:1975}.}
\label{fig:ff_o16}
\end{figure}

The charge form factors of \isotope[6]{Li}, \isotope[12]{C}, and \isotope[16]{O} are shown 
in~\cref{fig:ff_li6,fig:ff_c12,fig:ff_o16}, respectively. In all the plots, the blue (red) curve
is the AFDMC result for the N$^2$LO $E\mathbbm1$ potential ($E\tau$ for \isotope[12]{C}), with cutoff $R_0=1.0\,(1.2)\,\rm fm$.
Monte Carlo error bars are typically of the size of the lines within the momentum 
range considered here. Lighter shaded areas indicate the uncertainties from 
the truncation of the chiral expansion, according to~\cref{eq:err}.
Darker shaded areas are instead the theoretical error bands only considering the last
term of the prescription, i.e., taking into account the NLO and N$^2$LO results only.
AFDMC results are compared to experimental data and to available Monte
Carlo calculations employing the phenomenological potentials and one-body charge operators only.
No two-body operators are included in the calculation of the charge form factors in the 
current work. However, as shown in Refs.~\cite{Wiringa:1998,Lovato:2013,Mihaila:2000} for the three
different systems, such operators give a measurable contribution only 
for $q>2\,\rm fm^{-1}$, as they basically include relativistic corrections.

The charge form factor of \isotope[6]{Li} for the $E\mathbbm1$ interaction is compatible
with experimental data at low momentum for both cutoffs, with larger theoretical uncertainties
for the soft potential. Results for the $E\tau$ parametrization
show a similar behavior. The discrepancy for $q\gtrsim2\,\rm fm^{-1}$ is
due to the missing two-body currents. In fact, AFDMC results for local chiral forces 
are compatible with the VMC one-body results for AV18+UIX~\cite{Wiringa:1998}
up to high momentum.

A similar physical picture is given for both \isotope[12]{C} and \isotope[16]{O}, 
for which the positions of the first diffraction peaks in the form factors 
are well reproduced by the hard potentials within the theoretical error bands, 
and deviations from the experimental data occur at high momentum only.
For the soft $E\mathbbm1$ interaction instead, the description of the charge form factor in
\isotope[16]{O} is less accurate, with the position of the first diffraction 
peak overestimated, and the slope of $F_L(q)$ for $q=0$ underestimated, 
consistent with the smaller charge radius compared to the experimental value.
The difference with respect to the experimental results is however not as 
dramatic as for the soft $E\tau$ potential (see Ref.~\cite{Lonardoni:2017afdmc}),
and it is mostly covered by the very large theoretical error bands. These, 
in particular, are dominated by the LO contributions to the theoretical error
estimate, as shown by the difference between lighter and darker bands in the
form factor.

Finally, it is interesting to note that for all three systems, the local chiral interactions with the
hard cutoff $R_0=1.0\,\rm$ fm give the same physical description of the charge form factor as 
the phenomenological potentials, provided that one-body charge operators only are included in the calculation.

Two-body densities are related to the Coulomb sum rule, which is defined as the energy integral 
of the electromagnetic longitudinal response function. As with the charge form factor, the Coulomb sum rule
can be written as a ground-state expectation value~\cite{Mcvoy:1962}, leading to the relation:
\begin{align}
	S_L(q)=&\frac{1}{Z} \frac{1}{G_E^{p\,2}(Q_{\rm qe}^2)}\frac{1}{1+Q_{\rm qe}^2/(4 m_N^2)} \nonumber \\
    & \times\Big\{ G_E^{p\,2}(Q_{\rm qe}^2)\,\Big[\tilde{\rho}_{pp}(q)+Z\Big] \nonumber \\
    &        +G_E^{n\,2}(Q_{\rm qe}^2)\,\Big[\tilde{\rho}_{nn}(q)+(A-Z)\Big] \nonumber \\
    &        +2\,G_E^p(Q_{\rm qe}^2)\,G_E^n(Q_{\rm qe}^2)\,\tilde{\rho}_{np}(q) \nonumber \\ 
    & -\Big[G_E^p(Q_{\rm qe}^2)\,\tilde{\rho}_p(q)+G_E^n(Q_{\rm qe}^2)\,\tilde{\rho}_n(q)\Big]^2 \Big\}, 
	\label{eq:sl}
\end{align}
where $\tilde{\rho}_{\rm{NN}}(q)$ is the Fourier transform of the two-body point-nucleon densities defined in \cref{eq:rho_NN},
and $Q^2_{\rm qe}=\vb{q}^2-\omega^2_{\rm qe}$, with $\omega_{\rm qe}$ the energy transfer 
corresponding to the quasielastic peak.
Although the Coulomb sum rule is not directly an experimental observable (experimental information can be however
extracted from the longitudinal response function, as done in Ref.~\cite{Lovato:2016} for \isotope[12]{C}), it is
still an interesting quantity for the study of integral properties of the response of a nuclear many-body system to an
external probe.

\begin{figure}[b]
\includegraphics[width=\linewidth]{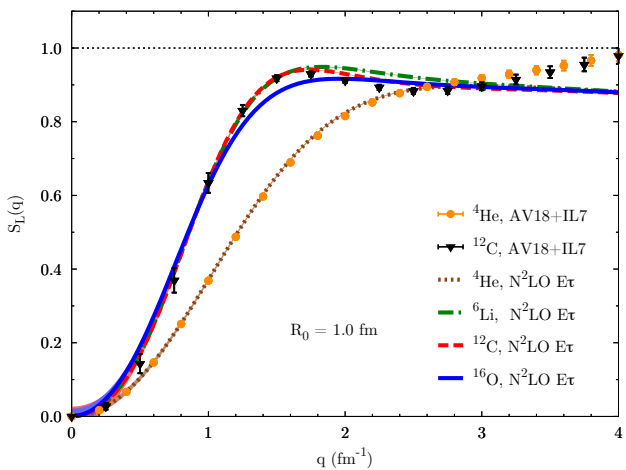}
\caption[]{Coulomb sum rule for $4\le A\le16$.
Lines refer to AFDMC results for the N$^2$LO $E\tau$ potential with cutoff $R_0=1.0\,\rm fm$.
Solid symbols are the GFMC one- plus two-body results for AV18+IL7~\cite{Lovato:2013,Lonardoni:2017cvmc}. 
Shaded areas indicate the statistical Monte Carlo uncertainty.}
\label{fig:sl}
\end{figure}

We report in~\cref{fig:sl} the Coulomb sum rule for $4\le A\le16$ using the N$^2$LO $E\tau$ potential
with cutoff $R_0=1.0\,\rm fm$. The GFMC results for \isotope[4]{He} 
and \isotope[12]{C}~\cite{Lovato:2013,Lonardoni:2017cvmc} employing the AV18+IL7 potential are also shown for comparison.
The discrepancy between the AFDMC and GFMC results above $\approx 3\,\rm fm^{-1}$ is due
to the missing two-body currents in the present calculation. For lower momenta the description of the sum rule 
is remarkably consistent with that provided by phenomenological potentials. 
Moreover, the results for \isotope[12]{C} are 
compatible with the available experimental data as extracted in Ref.~\cite{Lovato:2016}, 
as shown already in Ref.~\cite{Lonardoni:2017afdmc}. 
All $p$-shell nuclei show a similar profile for $S_L(q)$, with a peak around 
$1.6\,\rm fm^{-1}$, slightly more pronounced for open-shell systems $(A=6,12)$.
The same observations hold for the $E\mathbbm1$ parametrization of the $3N$ force 
and for both cutoffs, with the Coulomb sum rule 
of \isotope[4]{He} and \isotope[6]{Li} very close to those shown in~\cref{fig:sl}. 
An exception is the case of \isotope[16]{O}, for which $S_L(q)$
is largely different for the soft cutoff, consistent with the results for the charge form factor, 
as already shown in Ref.~\cite{Lonardoni:2017afdmc}.

\section{Summary}
\label{sec:summ}
We presented a detailed description of the AFDMC method for nuclei, 
with particular attention given to the construction of the trial wave function, the propagation 
of $3N$ forces, and the constrained/unconstrained imaginary-time evolution. 
We reported a series of test results for these technical aspects of the algorithm.

We performed AFDMC calculations of nuclei with $3\leq A\leq16$ using local chiral 
EFT interactions up to N$^2$LO, completing and expanding the 
results of Ref.~\cite{Lonardoni:2017afdmc}.
Both two- and three-body potentials have been considered, the latter described by
two different operator structures, namely $E\tau$ and $E\mathbbm1$. Two coordinate-space 
cutoffs, $R_0=1.0\,\rm fm$ and $R_0=1.2\,\rm fm$, have been used, with 
results presented at each order of the chiral expansion and for each $3N$ parametrization. 
To this aim, a new fit of the three-body LECs $c_D$ and $c_E$ has been presented for the 
$E\mathbbm1$ parametrization with the soft cutoff $R_0=1.2\,\rm fm$.

Binding energies and charge radii were shown for all the systems,
and results for the charge form factor in \isotope[6]{Li}, \isotope[12]{C}, and 
\isotope[16]{O} were also reported. For all these observables, the AFDMC results
were supported by statistical Monte Carlo errors and theoretical errors coming
from the truncation of the chiral expansion. Finally, the Coulomb sum rule for 
systems with $4\leq A\leq 16$ was also shown.

The outcomes of this work confirm that local chiral interactions fit to 
few-body observables give a very good description of the ground-state properties 
of nuclei up to \isotope[16]{O}. 
This is true for both harder and softer interactions, even though the latter 
imply larger theoretical uncertainties coming from LO contributions to
the truncation error estimate. We found that the overbinding in \isotope[16]{O}
for the soft $E\tau$ parametrization of the $3N$ force is generated by 
large attractive contributions from the large value of the LEC $c_D$. 
Therefore, it will be very interesting to explore further $3N$ fits and 
operator choices in heavier nuclei as well as dense matter.

\acknowledgments{We thank I.~Tews, A.~Lovato, A.~Roggero, and R.~F.~Garcia Ruiz 
for many valuable discussions. 
The work of D.L. was was supported by the U.S. Department of Energy, 
Office of Science, Office of Nuclear Physics, under the FRIB Theory 
Alliance Grant Contract No. DE-SC0013617 titled ``FRIB Theory Center 
- A path for the science at FRIB'', and by the NUCLEI SciDAC program.
The work of S.G. and J.C. was supported by the NUCLEI SciDAC program,
by the U.S. Department of Energy, Office of Science, Office of Nuclear
Physics, under contract No. DE-AC52-06NA25396, and by the LDRD program
at LANL.
K.E.S. was supported by the National Science Foundation grant PHY-1404405.
The work of J.E.L. and A.S. was supported by the ERC Grant No.~307986
STRONGINT and the BMBF under Contract No.~05P15RDFN1.
Computational resources have been provided by Los Alamos Open
Supercomputing via the Institutional Computing (IC) program, by the
National Energy Research Scientific Computing Center (NERSC), which is
supported by the U.S. Department of Energy, Office of Science, under
contract DE-AC02-05CH11231, and by the Lichtenberg high performance
computer of the TU Darmstadt.

\appendix
\section{Calculating two-body correlations}
Given $R=\{\vb{r}_1,\dots,\vb{r}_A\}$ the particle coordinates, 
$S=\{s_1,\dots,s_A\}$ the spin/isospin configurations, 
and $|\chi_\gamma\rangle$ the $|p\uparrow\rangle$, $|p\downarrow\rangle$,
$|n\uparrow\rangle$, $|n\downarrow\rangle$ basis:
\begin{align}
	|\chi_1\rangle=|(1,0,0,0)\rangle , \nonumber \\
	|\chi_2\rangle=|(0,1,0,0)\rangle , \nonumber \\
	|\chi_3\rangle=|(0,0,1,0)\rangle , \nonumber \\
	|\chi_4\rangle=|(0,0,0,1)\rangle ,
\end{align}
We define the Slater matrix element
\begin{align}
   S_{\alpha i}=\langle\alpha|\mathbf{r}_i\,s_i\rangle=\sum_{\gamma=1}^4\langle\alpha|\mathbf{r}_i\,\chi_\gamma\rangle\langle\chi_\gamma|s_i\rangle ,
\end{align}
where $|\alpha\rangle$ contains the radial orbitals and spherical harmonics of~\cref{eq:phi}. 
When acting with two-body correlations on the mean-field part of the wave function, 
the Slater matrix is updated by each of the correlation operators. 
These updates are computed using the identity 
\begin{align}
	\det\left(S^{-1}S'\right) = \frac{\det S'}{\det S} ,
	\label{eq:invers}
\end{align}
where $S'$ is the matrix that has been updated by the action of a single operator. 
To reduce the number of operations, the ratio of determinants for a pair of operators, 
$\mathcal{O}_{ij}=\mathcal{O}_i\mathcal{O}_j$, is written in the form
\begin{align}
   \frac{\langle \Phi|\mathcal O_{ij}|RS\rangle }{\langle\Phi|RS\rangle} = \sum_{\gamma=1}^4\sum_{\delta=1}^4 d_{2b}(\chi_\gamma,\chi_\delta,ij)\langle \chi_\gamma\chi_\delta|\mathcal O_{ij}|s_is_j\rangle ,
	\label{eq:ratio}
\end{align}
with
\begin{widetext}
\begin{align}
 d_{2b}(\chi_\gamma,\chi_\delta,ij)=\frac{\langle\Phi|R,s_1,\ldots,s_{i-1},\chi_\gamma,s_{i+1},\ldots,s_{j-1},\chi_\delta,s_{j+1},\ldots,s_A\rangle}{\langle \Phi|RS\rangle} ,
\end{align}
\end{widetext}
where $\chi_\gamma$ and $\chi_\delta$ replace $s_i$ and $s_j$, respectively.
The $d_{2b}$ matrix elements are derived from the precalculated matrix elements $P_{\chi,ij}$
\begin{align}
   d_{2b}(\chi_\gamma,\chi_\delta,ij)=\det\begin{pmatrix}P_{\chi_\gamma,ii} & P_{\chi_\gamma,ij} \\ P_{\chi_\delta,ji} & P_{\chi_\delta,jj}\end{pmatrix} ,
\end{align}
where
\begin{align}
   P_{\chi_\gamma,ij} &=\sum_\alpha S^{-1}_{j\alpha}S_{\alpha i}(s_i\leftarrow \chi_\gamma) , \nonumber \\
   P_{\chi_\delta,ij} &=\sum_\alpha S^{\prime\;-1}_{j\alpha}S^\prime_{\alpha i}(s_j\leftarrow \chi_\delta) .
\end{align}

Though the above relations only address two-body operators, 
this method can be generalized to arbitrary $N$-body operators as well. 
To include additional operators the matrix elements $P_{\chi,ij}$ need to be updated
\begin{align}
   P_{\chi_\eta,mn}=\sum_\alpha S^{\prime\prime\;-1}_{n\alpha}S^{\prime\prime}_{\alpha m}(s_m\leftarrow \chi_\eta) ,
\end{align}
where
\begin{align}
   S^{\prime\prime}_{\alpha m}(s_m) = \left\{
   \begin{array}{cc}
      S_{\alpha m} & m \ne i\\
      \langle\alpha|\mathcal O_i|\mathbf{r}_i\,s_i\rangle  & m = i
   \end{array} .
   \right.
\end{align}
To calculate the updated inverse matrix, the identity of~\cref{eq:invers} 
is used with $S^\prime\leftarrow S^{\prime\prime}$. 
Both sides of the identity are expanded, and like terms are grouped, 
noting that when $j \ne i$, $S^{\prime\prime}_{mi}=S^\prime_{mi}$.

The wave function with linear correlations (\cref{eq:psi}) is calculated 
by first acting on the coordinate and spin/isospin configurations with 
each possible operator, and calculating the sum of each term 
$\sum_{\chi_\gamma,\chi_\delta}d_{2b}(\chi_\gamma,\chi_\delta,ij)\langle\chi_\gamma,\chi_\delta|f_{ij}^p\mathcal{O}_{ij}^p|s_i s_j\rangle$. 
The expectation value of the potential on the linear wave function 
is calculated including correlation and potential operators, 
$\mathcal{O}^c_{ij}$ and $\mathcal{O}^p_{ij}$ respectively, 
organized in the form $(\mathbbm1+\mathcal{O}^c_{ij})\mathcal{O}^p_{kl}$,
which includes four potentially distinct operators. 
For this calculation the $P$ matrix is updated twice, 
once for $\mathcal{O}^c_i$ and once for $\mathcal{O}^c_j$, 
where $\mathcal{O}^c_{ij}=\mathcal{O}^c_i\mathcal{O}^c_j$ as before.  
The ratio of determinants is calculated following~\cref{eq:ratio},
using the updated distribution $d^{\prime\prime}_{2b}$.

The quadratic wave function includes the same correlation terms 
of the linear wave function plus a piece with two additional operators,
resulting in structures like $\mathbbm1+\mathcal{O}^c_{ij}+\mathcal{O}^c_{ij}\mathcal{O}^c_{kl}$. 
The operators up to linear terms are treated as above. 
The quadratic product of operators is handled in the same fashion 
as the expectation value of the potential acting on the linear wave function, i.e., 
the $P$ matrix is updated twice, once for $\mathcal{O}^c_i$ and once 
for $\mathcal{O}^c_j$, and the ratio of determinants is calculated with 
the updated distributions. It follows that the calculation of the correlation 
operators for the quadratic wave function requires $O(A^4)$ operations, compared
to $O(A^2)$ for the linear wave function.

The expectation value of the potential acting on the quadratic wave function 
requires the product of six operators $\mathcal{O}^c_{ij}\mathcal{O}^c_{kl}\mathcal{O}^p_{mn}$. 
As a result, a total of four updates are needed to calculate the quadratically correlated 
terms for the potential. After including the updated 
distributions for the $\mathcal{O}^c_{ij}$ operators, the same distributions are updated 
two more times for the $\mathcal{O}^c_{kl}$ terms. These quadratically updated distributions 
are then used to calculate the expectation value of the potential as before. 
It follows that the calculation of the expectation value of the potential acting 
on the quadratic wave function requires $O(A^6)$ operations, compared
to $O(A^4)$ for the linear wave function.

The two-body correlations of~\cref{eq:psi} have the same operator structure as the 
AV6$'$ potential. The Cartesian breakup of such structure generates 39 $\mathcal O^c_{ij}$ operators,  
$9\;\sigma_{\alpha i}\,\sigma_{\beta j}$, $3\;\tau_{\gamma i}\,\tau_{\gamma j}$,
and $27\;\sigma_{\alpha i}\,\sigma_{\beta j}\,\tau_{\gamma i}\,\tau_{\gamma j}$ operators. 
The number of operators can be reduced to 15 if, instead of Cartesian coordinates, 
one uses the pair distance $\vb{r}_{ij}$ and two orthogonal coordinates. 
This reduces the number of operators used in the spatially dependent part of the tensor term, $3\,\bm\sigma_i\cdot\hat{\vb{r}}_{ij}\,\bm\sigma_j\cdot\hat{\vb{r}}_{ij}$, from 9 to 3.

\end{document}